\documentclass[
 twocolumn,
 superscriptaddress,
 nofootinbib,
 amsmath,
 amssymb,
 aps,
]{revtex4-2}

\usepackage{graphicx} 
\usepackage{dcolumn} 
\usepackage{bm} 

\usepackage{hyperref} 

\usepackage{xcolor}
\allowdisplaybreaks
\usepackage{soul}

\usepackage{mathtools}
\DeclareMathOperator*{\argmin}{arg\,min}
\DeclareMathOperator*{\argmax}{arg\,max}
\usepackage{bbold}

\usepackage{xr}
\renewcommand{\theequation}{\arabic{section}.\arabic{equation}}
\makeatletter
\@addtoreset{equation}{section}
\makeatother

\begin{document}

\preprint{APS/123-QED}

\title{Information geometric bound on general chemical reaction networks}

\author{Tsuyoshi Mizohata}

\email{mizohata.tsuyoshi.t8@elms.hokudai.ac.jp}

\affiliation{
	Graduate School of Information Science and Technology,
	Hokkaido University, Sapporo, Hokkaido, 060-0814, Japan
}

\author{Tetsuya J. Kobayashi}

\email{tetsuya@sat.t.u-tokyo.ac.jp}

\affiliation{
	Institute of Industrial Science, The University of Tokyo, 4-6-1, Komaba, Meguro-ku, Tokyo 153-8505 Japan
}

%
%

\author{Louis-S. Bouchard}

\email{louis.bouchard@gmail.com}

\affiliation{
	Center for Quantum Science and Engineering, University of California, Los Angeles, CA 90095, USA
}

\affiliation{
	Department of Chemistry and Biochemistry, University of California, Los Angeles, CA 90095, USA
}

\affiliation{
	California NanoSystems Institute, University of California, Los Angeles, CA 90095, USA
}

\author{Hideyuki Miyahara}

\email{miyahara@ist.hokudai.ac.jp, hmiyahara512@gmail.com}

\thanks{corresponding author.}

\affiliation{
	Graduate School of Information Science and Technology,
	Hokkaido University, Sapporo, Hokkaido, 060-0814, Japan
}




\date{\today}

\begin{abstract}
	We investigate the dynamics of chemical reaction networks (CRNs) with the goal of deriving an upper bound on their reaction rates.
	This task is challenging due to the nonlinear nature and discrete structure inherent in CRNs.
	To address this, we employ an information geometric approach, using the natural gradient, to develop a nonlinear system that yields an upper bound for CRN dynamics.
	We validate our approach through numerical simulations, demonstrating faster convergence in a specific class of CRNs.
	This class is characterized by the number of chemicals, the maximum value of stoichiometric coefficients of the chemical reactions, and the number of reactions.
	We also compare our method to a conventional approach, showing that the latter cannot provide an upper bound on reaction rates of CRNs.
	While our study focuses on CRNs, the ubiquity of hypergraphs in fields from natural sciences to engineering suggests that our method may find broader applications, including in information science.
\end{abstract}


\maketitle


\section{Introduction} \label{main_sec_introduction_001_001}

Over the past three decades, extensive research has been dedicated to understanding stochastic and information thermodynamics, specifically focusing on bounds related to entropy production and various physical quantities~\cite{Seifert_001, Shiraishi_001, Kawai_001, Miyahara_001}.
This trajectory persists, with newer studies shedding light on thermodynamic uncertainty relations~\cite{Barato_001, Gingrich_001, Hasegawa_001, Vu_001, Vu_002} and establishing thermodynamic bounds on cross-correlations~\cite{Ohga_001, Vu_003, Dechant_001}.
Parallel to the work on physical systems, researchers have also explored bounds in non-physical realms such as biological systems.
For example, limits concerning population growth have been studied~\cite{Kobayashi_001, Miyahara_002, Sughiyama_001, Miyahara_003, Adachi_001, Hoshino_001}.

Recent studies have unveiled the geometric structure of chemical reaction networks (CRNs) and have also extended these concepts to the domain of general hypergraphs~\cite{Kobayashi_003, Kobayashi_004, Kobayashi_005}.
Concurrently, topological analyses on CRNs and hypergraphs have been performed~\cite{Okada_001, Hirono_001, Hirono_002, Hirono_003}.
Despite these advancements, the intrinsic nonlinearity in CRNs presents a significant challenge for elucidating specific properties, leaving gaps in our understanding.

Information geometry offers a framework that applies differential geometry to probability distributions, facilitating the exploration of their geometric structures~\cite{Amari_002, Amari_004}.
Among its significant contributions is the concept of the natural gradient (NG)~\cite{Amari_005}, which has demonstrated effectiveness in optimization problems, particularly in the realm of machine learning.
Additional studies have ventured into the acceleration of information gradient flows~\cite{Wang_001} and have investigated the biological significance of gradient flows~\cite{Horiguchi_001}.
Research has also extended to constraints involving rates of statistical divergences and mutual information~\cite{Karbowski_001}.
These diverse applications underline the versatility of information geometry, which we leverage in this study.

In the present work, we explore the upper bound on reaction rates in general CRNs using NG.
Initially, we present a geometrical description of CRN dynamics~\cite{Kobayashi_004, Kobayashi_005}.
Subsequently, we categorize CRNs according to the number of chemicals involved, the maximum coefficients in the reactions, and the total number of reactions.
Utilizing this classification, we formulate a nonlinear system that provides an upper bound on reaction rates for a given class of CRNs.
Through numerical simulations, we find that our system exhibits a steeper gradient, facilitating faster convergence.
Importantly, this fast convergence minimizes the Kullback-Leibler (KL) divergence to zero.
In contrast, conventional CRNs often maintain a nonzero KL divergence due to nontrivial equilibrium points.
We also note that conventional methods are insufficient for achieving these results, underscoring the uniqueness of our approach.

The remainder of this paper is structured as follows.
In Sec.~\ref{main_sec_CRNs_001_001}, we furnish an overview of CRNs.
Section~\ref{main_sec_difficulty_001_001} elucidates the challenges of establishing an upper bound on CRNs using Newton's method.
Section~\ref{main_sec_natural_gradient_001_001} is dedicated to explaining NG.
In Sec.~\ref{main_sec_upper_bound_001_001}, we introduce a dynamical system that serves as an upper bound for CRNs in a specified class.
Numerical simulations are presented in Sec.~\ref{main_sec_numerical_simulations_001_001}.
In Sec.~\ref{main_sec_discussions_001_001}, we provide discussions on our findings.
The paper concludes with Sec.~\ref{main_sec_conclusions_001_001}.

\section{Chemical reaction networks} \label{main_sec_CRNs_001_001}

In this section, the primary aim is to formulate the geometric representation of the dynamical equations governing CRNs, as delineated in~\cite{Kobayashi_004, Kobayashi_005}.
We commence by presenting the standard notation for hypergraphs and CRNs.
Subsequently, we elucidate the dynamics intrinsic to CRNs, as well as concepts of Legendre duality and detailed balance.
These elements are then combined to construct the geometric expression of CRN dynamics.

\subsection{Definition of CRNs}

We begin with a hypergraph $(\mathbb{V}, \mathbb{E})$ where $\mathbb{V} \coloneqq \{ \mathbb{v}_i \}_{i=1}^{N_\mathbb{v}}$ and $\mathbb{E} \coloneqq \{ \mathbb{e}_e \}_{e=1}^{N_\mathbb{e}}$ as a hypergraph provides a mathematical framework to describe a chemical reaction.
Suppose that a CRN of interest involves $N_\mathbb{X}$ chemicals, denoted as $\mathbb{X}_1, \mathbb{X}_2, \dots, \mathbb{X}_{N_\mathbb{X}}$.
In the case of a CRN, each hypervertex $\mathbb{v}_i$ is composed of a combination of chemicals $\mathbb{X}_1, \mathbb{X}_2, \dots, \mathbb{X}_{N_\mathbb{X}}$ and given by
\begin{align}
	\mathbb{v}_i \coloneqq \gamma_{1, i} \mathbb{X}_1 + \gamma_{2, i} \mathbb{X}_2 + \dots + \gamma_{N_\mathbb{X}, i} \mathbb{X}_{N_\mathbb{X}}. \label{main_eq_def_hyperedge_001_001}
\end{align}
Each hyperedge $\mathbb{e}_e$ corresponds to a chemical reaction and is defined by a directed pair of two hypervertices $\mathbb{e}_e \coloneqq (\mathbb{v}_e^+, \mathbb{v}_e^-)$, which can be expressed as
\begin{align}
	 & \alpha_{1, e} \mathbb{X}_1 + \alpha_{2, e} \mathbb{X}_2 + \dots + \alpha_{N_\mathbb{X}, e} \mathbb{X}_{N_\mathbb{X}} \nonumber                                                                \\
	 & \quad \xrightarrow{\mathbb{e}_e} \beta_{1, e} \mathbb{X}_1 + \beta_{2, e} \mathbb{X}_2 + \dots + \beta_{N_\mathbb{X}, e} \mathbb{X}_{N_\mathbb{X}}. \label{main_eq_chemical_reaction_001_001}
\end{align}
Here, $\mathbb{v}_e^\pm$ are chosen from $\{ \mathbb{v}_i \}_{i=1}^{N_\mathbb{v}}$, and in Eq.~\eqref{main_eq_chemical_reaction_001_001}, $\mathbb{v}_e^+ = \alpha_{1, e} \mathbb{X}_1 + \alpha_{2, e} \mathbb{X}_2 + \dots + \alpha_{N_\mathbb{X}, e} \mathbb{X}_{N_\mathbb{X}}$ and $\mathbb{v}_e^- = \beta_{1, e} \mathbb{X}_1 + \beta_{2, e} \mathbb{X}_2 + \dots + \beta_{N_\mathbb{X}, e} \mathbb{X}_{N_\mathbb{X}}$.
We also define the order of reaction as follows:
\begin{align}
	m & \coloneqq \max_{i, e} \{ \alpha_{i, e}, \beta_{i, e} \}. \label{main_eq_def_m_001_001}
\end{align}
To characterize CRNs, $m$ in Eq.~\eqref{main_eq_def_m_001_001} will play an important role.

When a CRN involves multiple chemical reactions, the description provided above may be inadequate.
To describe a complex CRN, the stoichiometric matrix plays a crucial role.
The stoichiometric matrix $S$ is defined as an $N_\mathbb{X} \times N_\mathbb{e}$ matrix and is given by
\begin{align}
	S & \coloneqq [\bm{s}_1, \bm{s}_2, \dots, \bm{s}_{N_\mathbb{e}}], \label{main_eq_def_S_001_001}
\end{align}
where, for $e = 1, 2, \dots, N_\mathbb{e}$,
\begin{align}
	\bm{s}_e & \coloneqq
	\begin{bmatrix}
		\beta_{1, e} - \alpha_{1, e} \\ \beta_{2, e} - \alpha_{2, e} \\
		\vdots                       \\
		\beta_{N_\mathbb{X}, e} - \alpha_{N_\mathbb{X}, e}
	\end{bmatrix}.
\end{align}
That is, the $(j, e)$-th element of $S$ is given by $s_{j, e} = \beta_{j, e} - \alpha_{j, e}$ for $j = 1, 2, \dots, N_\mathbb{X}$ and $e = 1, 2, \dots, N_\mathbb{e}$.
In general, when a CRN involves multiple chemical reactions, the stoichiometric matrix provides a concise representation of the relationships between the reactants and products.

The stoichiometric matrix $S$ is also expressed as $S = - \Gamma B$.
Here, $B \in \{1, 0, -1\}^{N_\mathbb{v} \times N_\mathbb{e}}$ is the incidence matrix whose $(i, e)$-th element is given for $i = 1, 2, \dots, N_\mathbb{v}$ and $e = 1, 2, \dots, N_\mathbb{e}$ by
\begin{align}
	b_{i, e} & \coloneqq
	\begin{cases}
		1  & (\text{$\mathbb{v}_i$ is the head of hyperedge $\mathbb{e}_e$: $\mathbb{v}_i = \mathbb{v}_e^+$}), \\
		-1 & (\text{$\mathbb{v}_i$ is the tail of hyperedge $\mathbb{e}_e$: $\mathbb{v}_i = \mathbb{v}_e^-$}), \\
		0  & (\text{otherwise}),
	\end{cases}
\end{align}
and $\Gamma \in \mathbb{Z}_{\ge 0}^{N_\mathbb{X} \times N_\mathbb{v}}$ is given by
\begin{align}
	\Gamma & \coloneqq [\bm{\gamma}_1, \bm{\gamma}_2, \dots, \bm{\gamma}_{N_\mathbb{v}}],
\end{align}
where, using $\gamma_{1, i}, \gamma_{2, i}, \dots, \gamma_{N_\mathbb{X}, i}$ in Eq.~\eqref{main_eq_def_hyperedge_001_001}, $\bm{\gamma}_i$ is defined as
\begin{align}
	\bm{\gamma}_i & \coloneqq [\gamma_{1, i}, \gamma_{2, i}, \dots, \gamma_{N_\mathbb{X}, i}]^\intercal,
\end{align}
for $i = 1, 2, \dots, N_\mathbb{v}$.
Having defined the necessary variables to describe CRNs, we will now derive the equation that characterizes the dynamics of CRNs in the remainder of this section.

\subsection{Dynamics of CRNs}

To analyze the dynamics of a CRN, we introduce fluxes associated with each hyperedge.
Let $j_e^+ (\bm{x})$ and $j_e^- (\bm{x})$ denote the currents from the head to the tail and from the tail to the head of hyperedge $\mathbb{e}_e$, respectively, where $\bm{x}$ is the chemical concentration vector.
We define $\bm{j}^+ (\bm{x}) \coloneqq [j_1^+ (\bm{x}), j_2^+ (\bm{x}), \dots, j_{N_\mathbb{e}}^+ (\bm{x})]^\intercal$ and $\bm{j}^- (\bm{x}) \coloneqq [j_1^- (\bm{x}), j_2^- (\bm{x}), \dots, j_{N_\mathbb{e}}^- (\bm{x})]^\intercal$.

The law of mass action is widely observed to hold for CRNs and is considered one of the fundamental characteristics that differentiate CRNs from nonchemical hypergraphs.
Based on this, we make the assumption of mass action kinetics for the forward and reverse reaction fluxes on hyperedge $\mathbb{e}_e$ in Eq.~\eqref{main_eq_chemical_reaction_001_001}:
\begin{align}
	j_e^\pm (\bm{x}) & = k_e^\pm \sum_{i=1}^{N_\mathbb{v}} b_{i, e}^\pm \prod_{j=1}^{N_\mathbb{X}} x_j^{\gamma_{j, i}}, \label{main_eq_flux_j_001_001}
\end{align}
where, for $i = 1, 2, \dots, N_\mathbb{X}$ and $e = 1, 2, \dots, N_\mathbb{e}$,
\begin{align}
	b_{i, e}^+ \coloneqq \max (b_{i, e}, 0), \\
	b_{i, e}^- \coloneqq - \min (b_{i, e}, 0),
\end{align}
and $k_e^\pm$ are the reaction rate coefficients for the forward and backward currents on $\mathbb{e}_e$.
Expressed in vector notation, Eq.~\eqref{main_eq_flux_j_001_001} can be written as
\begin{align}
	\bm{j}^\pm (\bm{x}) & = \bm{k}^\pm \circ (B^\pm)^\intercal {\bm{x}^\Gamma}^\intercal \\
	                    & = \bm{k}^\pm \circ {\bm{x}^{(\Gamma B^\pm)}}^\intercal,
\end{align}
where
\begin{align}
	B^+                       & \coloneqq \max (B, \mathbb{0}),                                                                                   \\
	B^-                       & \coloneqq - \min (B, \mathbb{0}),                                                                                 \\
	{\bm{x}^\Gamma}^\intercal & \coloneqq [\bm{x}^{\bm{\gamma}_1}, \bm{x}^{\bm{\gamma}_2}, \dots, \bm{x}^{\bm{\gamma}_{N_\mathbb{v}}}]^\intercal, \\
	\bm{x}^{\bm{\gamma}_i}    & \coloneqq \prod_{j=1}^{N_\mathbb{X}} x_j^{\gamma_{j, i}},                                                         \\
	\bm{k}^\pm                & \coloneqq [k_1^\pm, k_2^\pm, \dots, k_{N_\mathbb{e}}^\pm].
\end{align}
Here, $\mathbb{0}$ represents the zero matrix, which has the same size as matrix $B$.
The functions $\max(\cdot, \cdot)$ and $\min(\cdot, \cdot)$ are applied elementwise, meaning that for each element $[A]_{i,j}$ and $[B]_{i,j}$, we have $[\max(A, B)]_{i,j} = \max([A]_{i,j}, [B]_{i,j})$ and $[\min(A, B)]_{i,j} = \min([A]_{i,j}, [B]_{i,j})$, respectively.
The notation $[\cdot]_{i,j}$ represents the element located at the $i$-th row and $j$-th column.
Moreover, the symbol $\circ$ denotes the element-wise product, which is defined as follows:
\begin{align}
	\bm{x} \circ \bm{y} & \coloneqq
	\begin{bmatrix}
		x_1 y_1 \\
		x_2 y_2 \\
		\vdots  \\
		x_{N_\mathbb{X}} y_{N_\mathbb{X}}
	\end{bmatrix},
\end{align}
where $\bm{x} \coloneqq [x_1, x_2, \dots, x_{N_\mathbb{X}}]^\intercal$, $\bm{y} \coloneqq [y_1, y_2, \dots, y_{N_\mathbb{X}}]^\intercal$.

The chemical concentration vector $\bm{x}_t$ at time $t$ satisfies the chemical rate equation (CRE) given by~\cite{Beard_001, Feinberg_001, Rao_001}
\begin{align}
	\dot{\bm{x}}_t & = S \bm{j} (\bm{x}_t), \label{main_eq_x_dot_j_001_001}
\end{align}
where $\bm{j} (\bm{x}) \coloneqq \bm{j}^+ (\bm{x}) - \bm{j}^- (\bm{x})$.

\subsection{Legendre duality of fluxes and forces}

In the realm of physics, the relationship between fluxes and forces is commonly expressed through Legendre duality, a concept that describes how forces and fluxes are dual aspects of the same system.
Their product results in entropy production, denoted as \( \langle \bm{j}, \bm{f} \rangle \).
In the context of chemical thermodynamics, we define the force on a hyperedge \( \mathbb{e}_e \) in a manner consistent with entropy production:
\begin{align}
	f_e (\bm{x}) & \coloneqq \frac{1}{2} \ln \frac{j_e^+ (\bm{x})}{j_e^- (\bm{x})}, \label{main_eq_def_force_001_001}
\end{align}
for $e = 1, 2, \dots, N_\mathbb{e}$.
The corresponding vector form is given by Eq.~\eqref{main_eq_def_force_001_001}, denoted as $$\bm{f} (\bm{x}) \coloneqq [f_1 (\bm{x}), f_2 (\bm{x}), \dots, f_{N_\mathbb{e}} (\bm{x})]^\intercal,$$
can be expressed as
\begin{align}
	\bm{f} (\bm{x}) & = \frac{1}{2} \ln \frac{\bm{j}^+ (\bm{x})}{\bm{j}^- (\bm{x})}, \label{main_eq_force_vector_001_001}
\end{align}
where the division and the logarithmic function are computed elementwise.

We introduce a quantity called ``frenetic activity,'' particularly on hyperedge \( \mathbb{e}_e \), to describe the rate of change in the state of the system $\mathbb{e}_e$~\cite{Kobayashi_004,Kobayashi_005} as
\begin{align}
	\omega_e (\bm{x}) & \coloneqq 2 \sqrt{j_e^+ (\bm{x}) j_e^- (\bm{x})}, \label{main_eq_def_activity_001_001}
\end{align}
for $e = 1, 2, \dots, N_\mathbb{e}$.
The vector form of Eq.~\eqref{main_eq_def_activity_001_001}, denoted as $\bm{\omega} (\bm{x}) \coloneqq [\omega_1 (\bm{x}), \omega_2 (\bm{x}), \dots, \omega_{N_\mathbb{e}} (\bm{x})]^\intercal$,
can be expressed as
\begin{align}
	\bm{\omega} (\bm{x}) & = 2 \sqrt{\bm{j}^+ (\bm{x}) \circ \bm{j}^- (\bm{x})}. \label{main_eq_def_activity_vector_001_001}
\end{align}
Then, the following strictly convex smooth function $\Psi_{\bm{\omega} (\bm{x})}^* (\bm{f} (\bm{x}))$, which is called the dissipation function, establishes the Legendre duality between force $\bm{f} (\bm{x})$, Eq.~\eqref{main_eq_force_vector_001_001} and flux $\bm{j} (\bm{x})$, Eq.~\eqref{main_eq_def_activity_vector_001_001}:
\begin{align}
	\Psi_{\bm{\omega} (\bm{x})}^* (\bm{f} (\bm{x})) & \coloneqq \bm{\omega} (\bm{x})^\intercal [\cosh (\bm{f} (\bm{x})) - \bm{1}],
\end{align}
where
\begin{align}
	\cosh (\bm{f} (\bm{x})) & \coloneqq
	\begin{bmatrix}
		\cosh (f_1 (\bm{x})) \\
		\cosh (f_2 (\bm{x})) \\
		\vdots               \\
		\cosh (f_{N_\mathbb{e}} (\bm{x}))
	\end{bmatrix},                                                                 \\
	\bm{f} (\bm{x})         & \coloneqq [f_1 (\bm{x}), f_2 (\bm{x}), \dots, f_{N_\mathbb{e}} (\bm{x})]^\intercal, \\
	\bm{1}                  & \coloneqq [\underbrace{1, 1, \dots, 1}_{N_\mathbb{e}}]^\intercal.
\end{align}
As a result we have~\footnote{We have used the following notation: $\partial_{\bm{f}} \Psi_{\bm{\omega} (\bm{x})}^* (\bm{f} (\bm{x})) = \partial_{\bm{f}} \Psi_{\bm{\omega} (\bm{x})}^* (\bm{f})|_{\bm{f} = \bm{f} (\bm{x})}$.}
\begin{align}
	\bm{j} (\bm{x}) & = \partial_{\bm{f}} \Psi_{\bm{\omega} (\bm{x})}^* (\bm{f} (\bm{x})). \label{main_eq_j_f_001_001}
\end{align}
Note that
\begin{align}
	\partial_{\bm{f}} \Psi_{\bm{\omega} (\bm{x})}^* (\bm{f} (\bm{x})) & = \bm{\omega} (\bm{x}) \circ \sinh (\bm{f} (\bm{x})) \\
	                                                                  & =
	\begin{bmatrix}
		\omega_1 (\bm{x}) \sinh (f_1 (\bm{x})) \\
		\omega_2 (\bm{x}) \sinh (f_2 (\bm{x})) \\
		\vdots                                 \\
		\omega_{N_\mathbb{e}} (\bm{x}) \sinh (f_{N_\mathbb{e}} (\bm{x}))
	\end{bmatrix}. \label{main_eq_partial_Psi_001_001}
\end{align}
Combining Eqs.~\eqref{main_eq_x_dot_j_001_001} and \eqref{main_eq_j_f_001_001}, we get
\begin{align}
	\dot{\bm{x}}_t & = S \partial_{\bm{f}} \Psi_{\bm{\omega} (\bm{x}_t)}^* (\bm{f} (\bm{x}_t)). \label{main_eq_x_dot_f_001_001}
\end{align}
While Eq.~\eqref{main_eq_x_dot_f_001_001} is a well-defined differential equation, it lacks an explicit functional form for \( \bm{f} (\bm{x}) \), thus limiting its predictive capability.
The functional form of \( \bm{f} (\bm{x}) \) based on thermodynamics and kinetics will be elaborated in the subsequent subsection.

\subsection{Chemical reaction dynamics}

Until this point, the discussion has centered on the general description of dynamics on hypergraphs.
Going forward, the focus will be exclusively on CRNs.
In the realm of chemical thermodynamics, it is a common assumption to employ mass action kinetics to describe reaction rates.
Within this framework, a specific definition of force is accepted and widely used~\cite{Feinberg_001, Beard_001, Kobayashi_004, Kobayashi_005}:
\begin{align}
	\bm{f} (\bm{x}) & = - \frac{1}{2} \bigg( S^\intercal \ln \bm{x} - \ln \frac{\bm{k}^+}{\bm{k}^-} \bigg). \label{main_eq_force_CRN_001_001}
\end{align}
To clarify the geometric meaning of Eq.~\eqref{main_eq_force_CRN_001_001}, we introduce the Bregman divergence $\mathcal{D}_\phi (\bm{x} \| \bm{y})$ associated with potential $\phi (\cdot)$~\footnote{We have used the notation: $\partial_{\bm{x}} \phi (\bm{y}) = \partial_{\bm{x}} \phi (\bm{x}) |_{\bm{x} = \bm{y}}$.}:
\begin{align}
	\mathcal{D}_\phi (\bm{x} \| \bm{y}) & \coloneqq \phi (\bm{x}) - \phi (\bm{y}) - \langle \bm{x} - \bm{y}, \partial_{\bm{x}} \phi (\bm{y}) \rangle. \label{main_eq_def_Bregman_divergence_001_001}
\end{align}
The derivative of Eq.~\eqref{main_eq_def_Bregman_divergence_001_001} is given by
\begin{align}
	\partial_{\bm{x}} \mathcal{D}_\phi (\bm{x} \| \bm{y}) & = \partial_{\bm{x}} \phi (\bm{x}) - \partial_{\bm{x}} \phi (\bm{y}).
\end{align}
The KL divergence is Eq.~\eqref{main_eq_def_Bregman_divergence_001_001} with the following potential~\footnote{See Appendix~\ref{main_sec_derivation_KL_divergence_001_001} for detail.}:
\begin{align}
	\phi_\mathrm{KL} (\bm{x}) & \coloneqq \sum_{i=1}^{N_\mathbb{X}} x_i \ln x_i. \label{main_eq_potential_KL_001_001}
\end{align}
Then, the KL divergence is defined by $\mathcal{D}_{\phi_\mathrm{KL}} (\cdot \| \cdot) \coloneqq \mathcal{D}_\mathrm{KL} (\cdot \| \cdot)$ and it reads
\begin{align}
	\mathcal{D}_\mathrm{KL} (\bm{x} \| \bm{y}) & = \sum_{i=1}^{N_\mathbb{X}} x_i \ln \frac{x_i}{y_i} - \sum_{i=1}^{N_\mathbb{X}} x_i + \sum_{i=1}^{N_\mathbb{X}} y_i, \label{main_eq_def_KL_divergence_001_001}
\end{align}
and its derivative takes the following form:
\begin{align}
	\partial_{\bm{x}} \mathcal{D}_\mathrm{KL} (\bm{x} \| \bm{y}) & =
	\begin{bmatrix}
		\ln x_1 - \ln y_1 \\
		\ln x_2 - \ln y_2 \\
		\vdots            \\
		\ln x_{N_\mathbb{X}} - \ln y_{N_\mathbb{X}}
	\end{bmatrix}.
\end{align}
Then, Eq.~\eqref{main_eq_force_CRN_001_001} is rewritten as
\begin{align}
	\bm{f} (\bm{x}) & = - \frac{1}{2} S^\intercal \partial_{\bm{x}} \mathcal{D}_\mathrm{KL} (\bm{x} \| \hat{\bm{x}}) + \bm{f}_\mathrm{ne}. \label{main_eq_f_x_001_001}
\end{align}
The definition of $\hat{\bm{x}}$ will be given in the following subsection, and $\bm{f}_\mathrm{ne} \not \in \mathrm{Im} [S^\intercal]$ represents the nonequilibrium force incurred to the system~\cite{Kobayashi_003}.

Mass action kinetics also offers the following definitions of the flux and activity~\cite{Feinberg_001, Beard_001,Kobayashi_004,Kobayashi_005}:
\begin{align}
	\bm{j} (\bm{x}) & = (\bm{k}^+ \circ (B^+)^\intercal - \bm{k}^- \circ (B^-)^\intercal) {\bm{x}^\Gamma}^\intercal. \label{main_eq_flux_CRN_001_001}
\end{align}
Substituting Eq.~\eqref{main_eq_flux_CRN_001_001} into Eq.~\eqref{main_eq_def_activity_vector_001_001}, we also get the activity for CRNs:
\begin{align}
	\bm{\omega} (\bm{x}) & = 2 \sqrt{\bm{k}^+ \circ \bm{k}^-} \circ \bm{x}^{R^\intercal / 2}. \label{main_eq_activity_CRN_001_001}
\end{align}
where
\begin{align}
	R & \coloneqq \Gamma (B^+ + B^-). \label{main_eq_def_R_001_001}
\end{align}
In the remaining part of this section, we will present the geometric expression of the equation for CRNs.

\subsection{Geometric expression of an equilibrium CRE}

Up to this point, the discussion has centered on the geometric relationships that exist among the chemical concentration, potential, force, and flux in a CRN.
Subsequently, the CRE specified in Eq.~\eqref{main_eq_x_dot_j_001_001} can be reformulated into a geometric expression~\cite{Beard_001, Feinberg_001, Rao_001}.
To accomplish this, the detailed balance condition (DBC) must be taken into account.
The DBC, a criterion for the dynamic stability of a system at equilibrium, is described in the following section~\cite{Kobayashi_004,Kobayashi_005}:
\begin{align}
	\ln \frac{\bm{k}^+}{\bm{k}^-} & = S^\intercal \ln \bm{x}_\mathrm{eq}, \label{main_eq_chemical_DBC_001_001}
\end{align}
Here, \(\bm{x}_\mathrm{eq}\) represents the equilibrium chemical concentration vector, which is dependent on both the initial concentration vector \(\bm{x}_\mathrm{ini}\) and the specific CRE under consideration.
Additionally, if Eq.~\eqref{main_eq_chemical_DBC_001_001} is met, then \(\bm{f}_\mathrm{ne} = \bm{0}\).
Generally, at equilibrium, net fluxes cease (\(\bm{j} = \bm{0}\)), allowing us to define a set of equilibrium chemical concentration vectors as follows:
\begin{align}
	V_\mathrm{eq} & \coloneqq \{ \bm{x} > \bm{0} | \bm{j} (\bm{x}) = \bm{0} \}. \label{main_eq_def_set_equilibrium_states_001_001}
\end{align}
From Eq.~\eqref{main_eq_chemical_DBC_001_001}, Eq.~\eqref{main_eq_def_set_equilibrium_states_001_001} is transformed into
\begin{align}
	V_\mathrm{eq} & = \{ \bm{x} > \bm{0} | \exists \bm{\eta} \in \mathbb{R}^{| \mathrm{ker} (S^\intercal) |}, \ln \bm{x} = \ln \bm{x}_\mathrm{eq} + U \bm{\eta} \}.
\end{align}
where $U \coloneqq [u_1, u_2, \dots, u_{| \mathrm{ker} (S^\intercal) |}]$ and $\{ u_i \}_{i=1}^{| \mathrm{ker} (S^\intercal)|}$ are the bases of $\mathrm{ker} (S^\intercal)$.
We have introduced $\hat{\bm{x}}$ in Eq.~\eqref{main_eq_f_x_001_001}.
We here impose the following relation to $\hat{\bm{x}}$:
\begin{align}
	\hat{\bm{x}} \in V_\mathrm{eq}. \label{main_eq_def_x_hat_in_V_eq_001_001}
\end{align}
Then Eq.~\eqref{main_eq_f_x_001_001} describes dynamics of gradient flow to $V_\mathrm{eq}$.
Equation~\eqref{main_eq_def_x_hat_in_V_eq_001_001} is equivalently written as
\begin{align}
	\ln \frac{\bm{k}^+}{\bm{k}^-} & = S^\intercal \ln \hat{\bm{x}}. \label{main_eq_relationship_k_x_hat_001_001}
\end{align}
Note that using $\hat{\bm{x}}$ instead of $\bm{x}_\mathrm{eq}$ provides us with a generalized expression of the dynamical system.

Finally, we have arrived at the geometric expression of a CRE.
Namely, combining Eqs.~\eqref{main_eq_x_dot_f_001_001}, \eqref{main_eq_f_x_001_001}, \eqref{main_eq_activity_CRN_001_001}, and \eqref{main_eq_chemical_DBC_001_001}, we get~\footnote{We have used the following notation: $\partial_{\bm{x}} \mathcal{D}_\mathrm{KL} (\bm{x}_t \| \hat{\bm{x}}) = \partial_{\bm{x}} \mathcal{D}_\mathrm{KL} (\bm{x} \| \hat{\bm{x}}) |_{\bm{x} = \bm{x}_t}$.}
\begin{align}
	\dot{\bm{x}}_t & = S \partial_{\bm{f}} \Psi_{\bm{\omega} (\bm{x}_t)}^* \bigg( - \frac{1}{2} S^\intercal \partial_{\bm{x}} \mathcal{D}_\mathrm{KL} (\bm{x}_t \| \hat{\bm{x}}) \bigg), \label{main_eq_CRN_001_001}
\end{align}
where $\hat{\bm{x}} \in V_\mathrm{eq}$.
Note that in Eq.~\eqref{main_eq_CRN_001_001}, replacing $\hat{\bm{x}}$ with $\bm{x}_\mathrm{eq}$ does not affect the dynamics of CRNs because $S U \eta = \bm{0}$.

\section{Difficulty of constructing an upper bound on the reaction rates of CRNs} \label{main_sec_difficulty_001_001}

In this section, we briefly revisit Newton's method and present a counterexample illustrating its limitations in establishing an upper bound on the reaction rates of CRNs.

\subsection{Newton's method}

As stated in Sec.~\ref{main_sec_introduction_001_001}, the objective of this paper is to determine an upper bound on the reaction rates of CRNs.
One might assume that straightforward optimization methods could achieve this.
However, before discussing NG, we elucidate the challenges of using Newton's method~\cite{Hamming_001} as an optimization technique for this purpose.
While the gradient method is another elementary optimization technique, its indeterminate step size precludes its consideration in this study.
We now turn to a specific optimization problem:
\begin{align}
	\min_{\bm{x}} f (\bm{x}). \label{main_eq_problem_Newton_001_001}
\end{align}
Letting $\bm{x}_t$ be the state at the $t$-th iteration for $t \in \mathbb{Z}_{\ge 0}$, Newton's method for Eq.~\eqref{main_eq_problem_Newton_001_001} is given by
\begin{align}
	\bm{x}_{t + 1} & = \bm{x}_t - [\partial_{\bm{x}}^2 f (\bm{x}_t)]^{-1} \partial_{\bm{x}} f (\bm{x}_t). \label{main_eq_update_Newton_method_001_001}
\end{align}
In the case of CRNs, we have $f (\bm{x}) = \mathcal{D}_\phi (\bm{x} \| \hat{\bm{x}})$; then Eq.~\eqref{main_eq_update_Newton_method_001_001} reads
\begin{align}
	\bm{x}_{t + 1} & = \bm{x}_t - G_\phi^{-1} (\bm{x}_t) \partial_{\bm{x}} \mathcal{D}_\phi (\bm{x}_t \| \hat{\bm{x}}), \label{main_eq_update_Newton_method_001_002}
\end{align}
where $G_\phi$ is the Hessian of $\phi (\cdot)$.

\subsection{Counterexample}

\begin{figure}[t]
	\centering
	\includegraphics[scale=0.500]{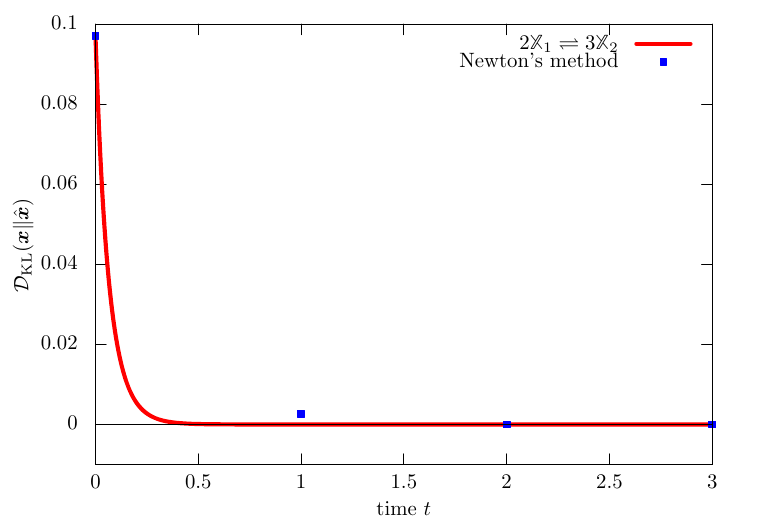}
	\caption{Dependence of $\mathcal{D}_\mathrm{KL} (\bm{x}_t \| \hat{\bm{x}})$ on $t$ for the CRN in Eq.~\eqref{main_eq_numerical_simulations_example_CRNs_002_001} and their upper bound in the case of $\bm{x}_\mathrm{eq} = \hat{\bm{x}}$.}
	\label{main_fig_upper_bound_not_eq_001_001}
\end{figure}

We will demonstrate a counterexample to show that Eq.~\eqref{main_eq_update_Newton_method_001_002} does not yield an upper bound for a CRN.
We consider the following CRN with $N_\mathbb{X} = 2$, $m = 3$, and $N_\mathbb{e} = 1$:
\begin{align}
	2 \mathbb{X}_1 & \rightleftharpoons 3 \mathbb{X}_2. \label{main_eq_numerical_simulations_example_CRNs_002_001}
\end{align}
For the simulations of Eq.~\eqref{main_eq_CRN_001_001}, we set $k_e^\pm = 1$, $\Delta t = 1.0 \times 10^{-4}$, $\bm{x}_\mathrm{ini} = [3/4, 11/8]^\intercal$, and $\hat{\bm{x}} = [1, 1]^\intercal$.
In Fig.~\ref{main_fig_upper_bound_not_eq_001_001}, we plot the dynamics of Eq.~\eqref{main_eq_numerical_simulations_example_CRNs_002_001} as well as the dynamics obtained using Newton's method.
At $t = 1$, the divergence of Newton's method is greater than that of the CRN, indicating that Newton's method fails to bound the dynamics.
This observation is illustrated in the figure.
The reason for this discrepancy lies in the nonlinearity of Eq.~\eqref{main_eq_numerical_simulations_example_CRNs_002_001}.

\section{Natural gradient} \label{main_sec_natural_gradient_001_001}

In this section, we explore the NG method and its applicability to the problem of constraining reaction rates in CRNs.
As our proposed methodology hinges on NG, understanding its theoretical underpinnings and its distinction from Newton's method is crucial.

\subsection{Derivation of NG}

In this section, we outline the derivation of the NG method, which is grounded in information geometry.
Specifically, we will elucidate how the dynamics of a given vector \(\bm{x}_t\) at time \(t\) are updated within the framework of NG:
\begin{align}
	\bm{x}_{t + \Delta t} & = \bm{x}_t + \Delta \bm{x}_t (\epsilon), \label{main_eq_NG_dynamics_001_001}
\end{align}
where $\Delta \bm{x}_t (\epsilon)$ is defined as~\footnote{We have used the following notation: $\partial_{\bm{x}} f (\bm{x}_t) = \partial_{\bm{x}} f (\bm{x}) |_{\bm{x} = \bm{x}_t}$.}
\begin{align}
	\Delta \bm{x}_t (\epsilon) & = \argmin_{\Delta \bm{x}: \mathcal{D}_{\phi'} (\bm{x}_t + \Delta \bm{x} \| \bm{x}_t) \le \epsilon} [f (\bm{x}_t + \Delta \bm{x}) - f (\bm{x}_t)]                                                                          \\
	                           & \approx \argmin_{\Delta \bm{x}: \frac{1}{2} \Delta \bm{x}^\intercal G_{\phi'} (\bm{x}_t) \Delta \bm{x} \le \epsilon} \partial_{\bm{x}} f (\bm{x}_t)^\intercal \Delta \bm{x}. \label{main_optimization_problem_NG_001_001}
\end{align}
Here, $G_{\phi'} (\bm{x}_t)$ is the Hessian given by
\begin{align}
	[G_{\phi'} (\bm{x}_t)]_{i, j} & \coloneqq \frac{\partial^2}{\partial x_i \partial x_j} \phi' (\bm{x}_t), \label{main_eq_def_Hessian_001_001}
\end{align}
where $[\cdot]_{i, j}$ is the $(i, j)$-th element.
In the case of Eq.~\eqref{main_eq_potential_KL_001_001}, Eq.~\eqref{main_eq_def_Hessian_001_001} reads
\begin{align}
	[G_{\phi'} (\bm{x}_t)]_{i, j} & = \delta_{i, j} \frac{1}{[\bm{x}_t]_i},
\end{align}
where $\delta_{i, j}$ is the Kronecker delta function and $[\cdot]_i$ is the $i$-th element.
To derive Eq.~\eqref{main_optimization_problem_NG_001_001}, we have used the following expansion of the Bregman divergence:
\begin{align}
	 & \mathcal{D}_{\phi'} (\bm{x}_t + \Delta \bm{x} \| \bm{x}_t) \nonumber                                                                                                          \\
	 & \quad = \phi' (\bm{x}_t + \Delta \bm{x}) - \phi' (\bm{x}_t) - \langle (\bm{x}_t + \Delta \bm{x}) - \bm{x}_t, \partial_{\bm{x}} \phi' (\bm{x}_t) \rangle                       \\
	 & \quad \approx \phi' (\bm{x}_t) + \partial_{\bm{x}} \phi (\bm{x}_t)^\intercal \Delta \bm{x} + \frac{1}{2} \Delta \bm{x}^\intercal G_{\phi'} (\bm{x}_t) \Delta \bm{x} \nonumber \\
	 & \qquad - \phi' (\bm{x}_t) - \langle (\bm{x}_t + \Delta \bm{x}) - \bm{x}_t, \partial_{\bm{x}} \phi' (\bm{x}_t) \rangle                                                         \\
	 & \quad = \frac{1}{2} \Delta \bm{x}^\intercal G_{\phi'} (\bm{x}_t) \Delta \bm{x}.
\end{align}
Note that $\Delta t$ in Eq.~\eqref{main_eq_NG_dynamics_001_001} is set to unity in the conventional formulation of NG; in the following section, we will impose a specific relationship between $\Delta t$ and $\epsilon$ in Eq.~\eqref{main_eq_NG_dynamics_001_001} to connect NG and CRNs.

To find the solution of Eq.~\eqref{main_optimization_problem_NG_001_001}, we employ the method of Lagrange multipliers where the Lagrange function reads
\begin{align}
	L (\Delta \bm{x}, \lambda) & \coloneqq \partial_{\bm{x}} f (\bm{x}_t)^\intercal \Delta \bm{x} - \frac{\lambda}{2} (\Delta \bm{x}^\intercal G_{\phi'} (\bm{x}_t) \Delta \bm{x} - \epsilon). \label{main_eq_Lagrange_function_NG_001_001}
\end{align}
The derivative of Eq.~\eqref{main_eq_Lagrange_function_NG_001_001} with respect to $\Delta \bm{x}$ takes the following form:
\begin{align}
	\frac{\partial}{\partial \Delta \bm{x}} L (\Delta \bm{x}, \lambda) & = \partial_{\bm{x}} f (\bm{x}_t) - \lambda G_{\phi'} (\bm{x}_t) \Delta \bm{x}. \label{main_eq_derivative_Lagrange_function_NG_001_001}
\end{align}
Then, the solution of Eq.~\eqref{main_eq_derivative_Lagrange_function_NG_001_001} is given by
\begin{align}
	\Delta \bm{x} & = \frac{1}{\lambda} G_{\phi'}^{-1} (\bm{x}_t) \partial_{\bm{x}} f (\bm{x}_t). \label{main_eq_solution_Delta_x_001_001}
\end{align}
The derivative of Eq.~\eqref{main_eq_Lagrange_function_NG_001_001} with respect to $\lambda$ has the following form:
\begin{align}
	\frac{\partial}{\partial \lambda} L (\Delta \bm{x}, \lambda) & = - \frac{1}{2} (\Delta \bm{x}^\intercal G_{\phi'} (\bm{x}_t) \Delta \bm{x} - \epsilon) \\
	                                                             & = 0. \label{main_eq_derivative_Lagrange_function_NG_002_001}
\end{align}
Taking Eq.~\eqref{main_eq_solution_Delta_x_001_001} into account, the solution of Eq.~\eqref{main_eq_derivative_Lagrange_function_NG_002_001} is written as
\begin{align}
	\lambda^2 & = \frac{\partial_{\bm{x}} f (\bm{x}_t)^\intercal G_{\phi'}^{-1} (\bm{x}_t) \partial_{\bm{x}} f (\bm{x}_t)}{\epsilon}. \label{main_eq_solution_lambda_001_001}
\end{align}
Combining Eqs.~\eqref{main_eq_solution_Delta_x_001_001} and \eqref{main_eq_solution_lambda_001_001} and taking account of the nature of the minimization problems, the solution of Eq.~\eqref{main_optimization_problem_NG_001_001} takes the following form:
\begin{align}
	\Delta \bm{x}_t (\epsilon) & = - \sqrt{\frac{\epsilon}{\partial_{\bm{x}} f (\bm{x}_t)^\intercal G_{\phi'}^{-1} (\bm{x}_t) \partial_{\bm{x}} f (\bm{x}_t)}} G_{\phi'}^{-1} (\bm{x}_t) \partial_{\bm{x}} f (\bm{x}_t).\label{main_eq_solution_NG_001_001}
\end{align}
Note that $\phi' (\cdot)$ in Eq.~\eqref{main_eq_solution_NG_001_001} may be different from $\phi (\cdot)$ appearing in Sec.~\ref{main_sec_CRNs_001_001}.
In the case of CRNs, $f (\bm{x}_t)$ in Eq.~\eqref{main_eq_solution_NG_001_001} represents $\mathcal{D}_\mathrm{KL} (\bm{x}_t \| \hat{\bm{x}})$.
As shown in Eq.~\eqref{main_eq_solution_NG_001_001}, $\epsilon$ is a key parameter in NG.
From the perspective of applying NG to CRNs, the relationship between $\epsilon$ in NG and $\Delta t$ in CRNs, when discretized, is still missing.
Therefore, NG cannot be directly applied to CRNs.
In the following section, we will explain how to address this challenge and develop a general upper bound on the dynamics of CRNs.

\subsection{Comparison with Newton's method}

In this section, we compare NG with Newton's method.
Newton's method is a special case of NG when Eq.~\eqref{main_eq_solution_NG_001_001} is adjusted according to certain conditions.
Specifically, the conditions are \(\phi (\cdot) = \phi' (\cdot)\) and \(\epsilon = \partial_{\bm{x}} f (\bm{x}_t)^\intercal G_{\phi'}^{-1} (\bm{x}_t) \partial_{\bm{x}} f (\bm{x}_t)\).
The equation thus becomes equivalent to Eq.~\eqref{main_eq_update_Newton_method_001_002}.
This equivalency leads us to introduce a systematic NG-based method to determine the direction and step size for a gradient system that bounds CRNs of a specific class.

\section{Upper bound on reaction rates} \label{main_sec_upper_bound_001_001}

In this section, we construct a nonlinear system that gives an upper bound on reaction rates of CRNs in a given class.
The class is characterized by several topological numbers of CRNs: $N_\mathbb{v}$, $N_\mathbb{e}$, and $m$.

\subsection{Upper bound system}

Comparing discretized CRE dynamics with NG dynamics, represented by Eq.~\eqref{main_eq_NG_dynamics_001_001}, presents a challenge.
The difficulty arises from the absence of an established relationship between \(\epsilon\), the constraint parameter in NG, and \(\Delta t\), the time step in the discretized CRE.
To address this issue, we propose the following relationship between \(\epsilon\) and \(\Delta t\):
\begin{align}
	\epsilon & = \mathcal{D}_{\phi'} (\bm{x}_t + \| \dot{\bm{x}}_t \|_\mathrm{F} \bm{e}_t \Delta t \| \bm{x}_t), \label{main_eq_epsilon_condition_001_001}
\end{align}
where $\| \cdot \|_\mathrm{F}$ is the Frobenius norm and $\bm{e}_t$ is a vector that satisfies $\| \bm{e}_t \|_\mathrm{F} = 1$.
Then, we try to compute the maximum value of $\epsilon$ in Eq.~\eqref{main_eq_epsilon_condition_001_001}.
Note that $S: \mathbb{R}^{N_\mathbb{e}} \to \mathbb{R}^{N_\mathbb{X}}$ and $S$ is a $N_\mathbb{X} \times N_\mathbb{e}$ matrix.
From Eq.~\eqref{main_eq_CRN_001_001}, we get
\begin{align}
	 & \| \dot{\bm{x}}_t \|_\mathrm{F} \nonumber                                                                                                                                                                                                                                                                             \\
	 & \quad = \bigg\| S \partial_{\bm{f}} \Psi_{\bm{\omega} (\bm{x}_t)}^* \bigg( - \frac{1}{2} S^\intercal \partial_{\bm{x}} \mathcal{D}_\phi (\bm{x}_t \| \hat{\bm{x}}) \bigg) \bigg\|_\mathrm{F}                                                                                                                          \\
	 & \quad \le \| S \|_\mathrm{F} \bigg\| \partial_{\bm{f}} \Psi_{\bm{\omega} (\bm{x}_t)}^* \bigg( - \frac{1}{2} S^\intercal \partial_{\bm{x}} \mathcal{D}_\phi (\bm{x}_t \| \hat{\bm{x}}) \bigg) \bigg\|_\mathrm{F}                                                                                                       \\
	 & \quad \le \| S \|_\mathrm{F} \bigg\| \partial_{\bm{f}} \Psi_{\bm{\omega} (\bm{x}_t)}^* \bigg( \bigg\| - \frac{1}{2} S^\intercal \partial_{\bm{x}} \mathcal{D}_\phi (\bm{x}_t \| \hat{\bm{x}}) \bigg\|_\mathrm{abs} \bigg) \bigg\|_\mathrm{F}                                                                          \\
	 & \quad \le \| S \|_\mathrm{F} \bigg\| \partial_{\bm{f}} \Psi_{\bm{\omega} (\bm{x}_t)}^* \bigg( \frac{1}{2} \|S^\intercal \|_\mathrm{F} \| \partial_{\bm{x}} \mathcal{D}_\phi (\bm{x}_t \| \hat{\bm{x}}) \|_\mathrm{F}^{N_\mathbb{X} \to N_\mathbb{e}} \bigg) \bigg\|_\mathrm{F}. \label{main_eq_maximum_x_dot_001_001}
\end{align}
Here, $\| \cdot \|_\mathrm{abs}$ and $\| \cdot \|_\mathrm{F}^{N_\mathbb{X} \to N_\mathbb{e}}$ are defined as, respectively,
\begin{align}
	\| \bm{v} \|_\mathrm{abs}                               & \coloneqq [| v_1 |, | v_2 |, \dots, | v_{N_\mathbb{X}} |]^\intercal,                                                                \\
	\| \bm{v} \|_\mathrm{F}^{N_\mathbb{X} \to N_\mathbb{e}} & \coloneqq [\underbrace{\| \bm{v} \|_\mathrm{F}, \| \bm{v} \|_\mathrm{F}, \dots, \| \bm{v} \|_\mathrm{F}}_{N_\mathbb{e}}]^\intercal,
\end{align}
for $\bm{v} \coloneqq [v_1, v_2, \dots, v_{N_\mathbb{X}}]^\intercal$.
From Eq.~\eqref{main_eq_partial_Psi_001_001}, we have
\begin{align}
	\partial_{\bm{f}} \Psi_{\bm{\omega} (\bm{x})}^* (\| \bm{f} (\bm{x}) \|_\mathrm{abs})                               & = \bm{\omega} (\bm{x}) \circ \sinh (\| \bm{f} (\bm{x}) \|_\mathrm{abs}),                               \\
	\partial_{\bm{f}} \Psi_{\bm{\omega} (\bm{x})}^* (\| \bm{f} (\bm{x}) \|_\mathrm{F}^{N_\mathbb{X} \to N_\mathbb{e}}) & = \bm{\omega} (\bm{x}) \circ \sinh (\| \bm{f} (\bm{x}) \|_\mathrm{F}^{N_\mathbb{X} \to N_\mathbb{e}}).
\end{align}
Given $S: \mathbb{R}^{N_\mathbb{e}} \to \mathbb{R}^{N_\mathbb{X}}$ and $\bm{v} \in \mathbb{R}^{N_\mathbb{X}}$, we have the following inequality for $e = 1, 2, \dots, N_\mathbb{e}$:
\begin{align}
	[\| S^\intercal \bm{v} \|_\mathrm{abs}]_e & \le \| S^\intercal \|_\mathrm{F} \| \bm{v} \|_\mathrm{F}                                      \\
	                                          & = [ \| S^\intercal \|_\mathrm{F} \| \bm{v} \|_\mathrm{F}^{N_\mathbb{X} \to N_\mathbb{e}} ]_e,
\end{align}
where $[\cdot]_e$ is the $e$-th element.
Then, we have finished computing the bound on $\| \dot{\bm{x}}_t \|_\mathrm{F}$ within a given class of CRNs.

Next, we compute $\bm{e}_t$ as follows:
\begin{align}
	\bm{e}_t & = \argmax_{\bm{e}: \| \bm{e} \|_\mathrm{F} = 1} \mathcal{D}_{\phi'} (\bm{x}_t + \| \dot{\bm{x}}_t \|_\mathrm{F} \bm{e} \Delta t \| \bm{x}_t)                                \\
	         & \approx \argmax_{\bm{e}: \| \bm{e} \|_\mathrm{F} = 1} \bigg( \frac{1}{2} \| \dot{\bm{x}}_t \|_\mathrm{F}^2 (\Delta t)^2 \bm{e}^\intercal G_{\phi'} (\bm{x}_t) \bm{e} \bigg) \\
	         & = \argmax_{\bm{e}: \| \bm{e} \|_\mathrm{F} = 1} \bm{e}^\intercal G_{\phi'} (\bm{x}_t) \bm{e}. \label{main_eq_optimal_e_vector_001_001}
\end{align}
Thus, $\bm{e}_t$ is the eigenvector associated with the maximum eigenvalue of $G_{\phi'} (\bm{x}_t)$.
Substituting Eq.~\eqref{main_eq_maximum_x_dot_001_001} and the solution of Eq.~\eqref{main_eq_optimal_e_vector_001_001} into Eq.~\eqref{main_eq_epsilon_condition_001_001}, we can calculate the maximum value of $\epsilon$ within a given class of CRNs.

\subsection{$S$ and $R$ of an upper bound system}

To identify the upper bound described by Eq.~\eqref{main_eq_maximum_x_dot_001_001} for CRNs under certain constraints, both \( S \) in Eq.~\eqref{main_eq_def_S_001_001} and \( R \) in Eq.~\eqref{main_eq_def_R_001_001} must be carefully designed.
We introduce a method for determining \( S_\mathrm{ub} \) and \( R_\mathrm{ub} \) specific to a class of CRNs characterized by \( N_\mathbb{X} \) as the number of chemicals, \( m \) as the highest coefficient in chemical reactions, and \( N_\mathbb{e} \) as the number of reactions.
The \( S_\mathrm{ub} \) and \( R_\mathrm{ub} \) matrices are of dimensions \( N_\mathbb{X} \times N_\mathbb{e} \), and their elements at the \( (i, e) \)-th position are defined as follows:
\begin{align}
	[S_\mathrm{ub}]_{i, e} & \coloneqq m, \label{main_eq_S_ub_001}                                                                                     \\
	[R_\mathrm{ub}]_{i, e} & \coloneqq \mathbb{1} [x_i \le 1] \min_i ([R]_{i, e}) + \mathbb{1} [x_i > 1] \max_i ([R]_{i, e}). \label{main_eq_R_ub_001}
\end{align}
Here, $\mathbb{1} [\cdot]$ denotes the indicator function, and $[\cdot]_{i, e}$ represents the $(i, e)$-th element.
The reader may think that $\mathbb{1} [\cdot]$ is not necessary.
This reflects the fact that $x^n \ge x^m$ for $x \in [1, \infty)$ and $n \ge m$ but $x^n \le x^m$ for $x \in (0, 1]$ and $n \ge m$.
By solving Eq.~\eqref{main_eq_solution_NG_001_001} with Eqs.~\eqref{main_eq_epsilon_condition_001_001}, \eqref{main_eq_maximum_x_dot_001_001}, \eqref{main_eq_optimal_e_vector_001_001}, \eqref{main_eq_S_ub_001}, and \eqref{main_eq_R_ub_001}, we can compute the upper bound for a given class.
In other words, we use the following inequality to construct an upper bound system:
\begin{align}
	 & \| \dot{\bm{x}}_t \|_\mathrm{F} \nonumber                                                                                                                                                                                                                                                \\
	 & \quad \le \| S_\mathrm{ub} \|_\mathrm{F} \nonumber                                                                                                                                                                                                                                       \\
	 & \qquad \times \bigg\| \partial_{\bm{f}} \Psi_{\bm{\omega}_\mathrm{ub} (\bm{x}_t)}^* \bigg( \frac{1}{2} \|S_\mathrm{ub}^\intercal \|_\mathrm{F} \| \partial_{\bm{x}} \mathcal{D}_\phi (\bm{x}_t \| \hat{\bm{x}}) \|_\mathrm{F}^{N_\mathbb{X} \to N_\mathbb{e}} \bigg) \bigg\|_\mathrm{F},
\end{align}
where
\begin{align}
	\bm{\omega}_\mathrm{ub} (\bm{x}) & \coloneqq 2 \sqrt{\bm{k}^+ \circ \bm{k}^-} \circ \bm{x}^{R_\mathrm{ub}^\intercal / 2}.
\end{align}

\subsection{Upper bound system with the KL constraint}

We utilize Eq.~\eqref{main_eq_potential_KL_001_001}, represented as $\phi' (\cdot) = \phi_\mathrm{KL} (\cdot)$, as the potential function for the Bregman divergence in the constraint of NG~\footnote{While there are many different candidates for $\phi' (\cdot)$, the $L^2$ constraint is often used.
	Then, we explain the case of the $L^2$ constraint in Appendix~\ref{main_sec_NG_L2_constraint_001_001}.}.
Subsequently, by substituting $\| \partial_{\bm{x}} \mathcal{D}_\mathrm{KL} (\bm{x}_t \| \hat{\bm{x}}) \|_\mathrm{F}^{N_\mathbb{X} \to N_\mathbb{e}}$ into Eq.~\eqref{main_eq_maximum_x_dot_001_001}, we can determine the maximum value of $\| \dot{\bm{x}}_t \|_\mathrm{F}$ as stated in Eq.~\eqref{main_eq_maximum_x_dot_001_001}.

\section{Numerical simulations} \label{main_sec_numerical_simulations_001_001}

In this section, numerical simulations are conducted to elucidate the upper-bound dynamics for a specified class of CRNs.
The parameters are set as follows: \( N_\mathbb{X} = 4 \), \( m = 4 \), and \( N_\mathbb{e} = 1 \).
The initial condition is chosen as \( \hat{\bm{x}} = [1, 1, 1, 1]^\intercal \) and the time step as \( \Delta t = 1.0 \times 10^{-5} \).
The rate constants \( k_e^\pm \) are fixed at 1 for all \( e \) ranging from 1 to \( N_\mathbb{e} \).
Simulations are executed for a total of \( 3.0 \times 10^4 \) steps.
The initial chemical concentration vector at time \( t = 0 \) is denoted as \( \bm{x}_\mathrm{ini} \).

\subsection{Case where CRNs have different equilibrium state}

We introduce CRNs that satisfy the same conditions and compare them from the viewpoint of reaction rate.
Here we consider the following six different reactions, which have the same topological quantities ($N_\mathbb{X} = 4$, $m = 4$, and $N_\mathbb{e} = 1$):
\begin{subequations} \label{main_eq_numerical_simulations_examples_CRNs_001_001}
	\begin{align}
		\mathbb{X}_1 + 4 \mathbb{X}_2   & \rightleftharpoons 4 \mathbb{X}_3 + 4 \mathbb{X}_4,                                                             \\
		4 \mathbb{X}_1 + 4 \mathbb{X}_2 & \rightleftharpoons 4 \mathbb{X}_3 + 4 \mathbb{X}_4,  \label{main_eq_numerical_simulations_example_CRNs_001_002} \\
		\mathbb{X}_1 + 2 \mathbb{X}_2   & \rightleftharpoons \mathbb{X}_3 + 3 \mathbb{X}_4,                                                               \\
		4 \mathbb{X}_1                  & \rightleftharpoons 4 \mathbb{X}_2 + 4 \mathbb{X}_3 + 4 \mathbb{X}_4,                                            \\
		\mathbb{X}_1                    & \rightleftharpoons 2 \mathbb{X}_2 + 2 \mathbb{X}_3 + 3 \mathbb{X}_4,                                            \\
		2 \mathbb{X}_1                  & \rightleftharpoons 3 \mathbb{X}_2 + 2 \mathbb{X}_3 + 3 \mathbb{X}_4.
	\end{align}
\end{subequations}

\begin{figure}[t ]
	\centering
	\includegraphics[scale=0.500]{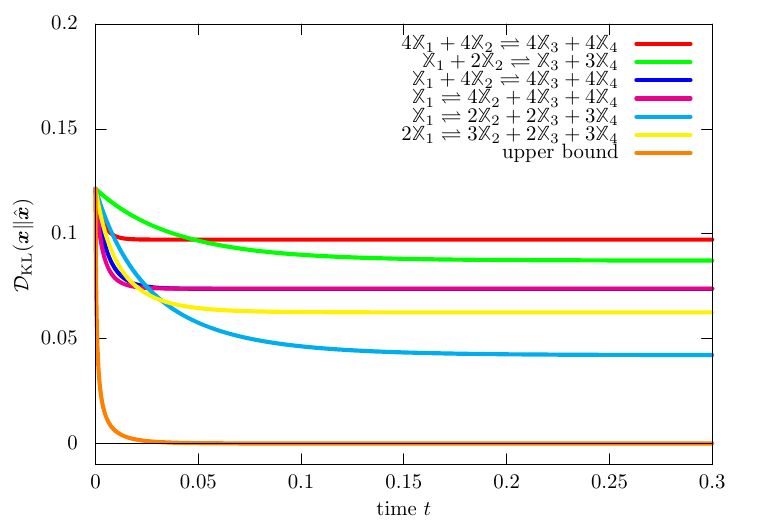}
	\caption{Dependence of $\mathcal{D}_\mathrm{KL} (\bm{x}_t \| \hat{\bm{x}})$ on time $t$ for several CRNs in Eq.~\eqref{main_eq_numerical_simulations_examples_CRNs_001_001} and their upper bound in the case of $\bm{x}_\mathrm{eq} \ne \hat{\bm{x}}$.}
	\label{main_fig_upper_bound_not_eq_002_001}
\end{figure}

We set $\bm{x}_\mathrm{ini} = [9/8, 87/80, 27/20, 27/20]^\intercal$, $\hat{\bm{x}} = [1, 1, 1, 1]^\intercal$, and $\Delta t = 1.0 \times 10^{-5}$.
In Fig.~\ref{main_fig_upper_bound_not_eq_002_001}, we plot the dynamics of Eq.~\eqref{main_eq_numerical_simulations_examples_CRNs_001_001} and that of the system constructed in Sec.~\ref{main_sec_upper_bound_001_001}.
It clearly shows that the system constructed in Sec.~\ref{main_sec_upper_bound_001_001} gives an upper bound on CRNs.
The CRNs in Eq.~\eqref{main_eq_numerical_simulations_examples_CRNs_001_001} have equilibrium states different from $\hat{\bm{x}}$ because of $\mathrm{ker} (S^\intercal)$; then the gap in $\mathcal{D}_\mathrm{KL} (\bm{x}_t \| \hat{\bm{x}})$ remains for $t \gg 0$ and the upper bound is relatively loose.

\subsection{Case where CRNs do not have different equilibrium state}

\begin{figure}[t]
	\centering
	\includegraphics[scale=0.500]{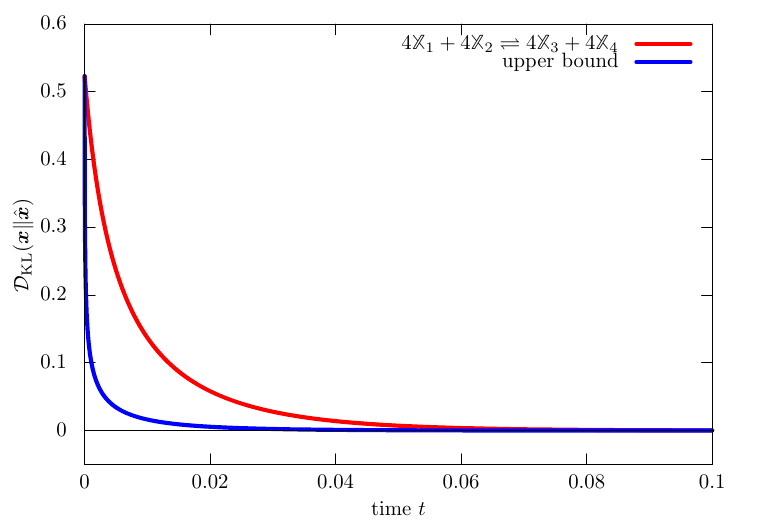}
	\caption{Dependence of $\mathcal{D}_\mathrm{KL} (\bm{x}_t \| \hat{\bm{x}})$ on time $t$ for the CRN in Eq.~\eqref{main_eq_numerical_simulations_example_CRNs_001_002} and its upper bound in the case of $\bm{x}_\mathrm{eq} = \hat{\bm{x}}$.}
	\label{main_fig_upper_bound_not_eq_003_001}
\end{figure}

\begin{figure}[t]
	\centering
	\includegraphics[scale=0.500]{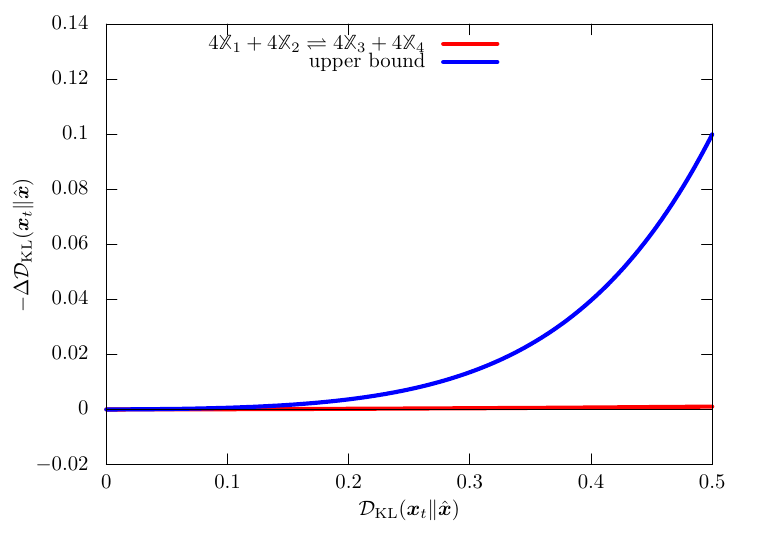}
	\caption{Relationship between $\mathcal{D}_\mathrm{KL} (\bm{x}_t \| \hat{\bm{x}})$ and $- \Delta \mathcal{D}_\mathrm{KL} (\bm{x}_t \| \hat{\bm{x}})$ for the CRN in Eq.~\eqref{main_eq_numerical_simulations_example_CRNs_001_002} and its upper bound in the case of $\bm{x}_\mathrm{eq} = \hat{\bm{x}}$. We have used $\bm{x}_t$ on the solution of Eq.~\eqref{main_eq_numerical_simulations_example_CRNs_001_002}.}
	\label{main_fig_upper_bound_not_eq_003_002}
\end{figure}

Next, we consider Eq.~\eqref{main_eq_numerical_simulations_example_CRNs_001_002} and set $\bm{x}_\mathrm{ini} = [1/2, 1/2, 3/2, 3/2]^\intercal$, $\hat{\bm{x}} = [1, 1, 1, 1]^\intercal$, and $\Delta t = 1.0 \times 10^{-5}$.
In this case, we have $\bm{x}_\mathrm{eq} = \hat{\bm{x}}$.
In Fig.~\ref{main_fig_upper_bound_not_eq_003_001}, we plot the dynamics of Eq.~\eqref{main_eq_numerical_simulations_examples_CRNs_001_001} and that of the system constructed in Sec.~\ref{main_sec_upper_bound_001_001}.
The system constructed in Sec.~\ref{main_sec_upper_bound_001_001} provides a tighter bound.
In Fig.~\ref{main_fig_upper_bound_not_eq_003_002}, we show the time-difference of the KL divergence $- \Delta \mathcal{D}_\mathrm{KL} (\bm{x}_t \| \hat{\bm{x}})$ per $\Delta t$.
We have used $\bm{x}_t$ on the solution of Eq.~\eqref{main_eq_numerical_simulations_example_CRNs_001_002} with $\bm{x}_\mathrm{ini} = [1/2, 1/2, 3/2, 3/2]^\intercal$; that is, $-\Delta \mathcal{D}_\mathrm{KL} (\bm{x}_t \| \hat{\bm{x}})$ of the CRN in Eq.~\eqref{main_eq_numerical_simulations_example_CRNs_001_002} and the system constructed in Sec.~\ref{main_sec_upper_bound_001_001} on the orbit of the CRN in Eq.~\eqref{main_eq_numerical_simulations_example_CRNs_001_002}.
As shown in Fig.~\ref{main_fig_upper_bound_not_eq_003_002}, the system constructed in Sec.~\ref{main_sec_upper_bound_001_001} shows faster convergence at each $\bm{x}_t$.

\subsection{Case of $N_\mathbb{e} > 1$}

We consider the fully-connected CRNs whose hypervertices are given by
\begin{subequations} \label{main_eq_numerical_simulations_example_CRNs_004_001}
	\begin{align}
		\mathbb{V}_1 & = \{ \mathbb{X}_1 + \mathbb{X}_2, \mathbb{X}_2 + \mathbb{X}_3, \mathbb{X}_3 + \mathbb{X}_4, \mathbb{X}_4 + \mathbb{X}_1 \},                                            \\
		\mathbb{V}_2 & = \{ \mathbb{X}_1 + 3 \mathbb{X}_2 + 4 \mathbb{X}_3, \mathbb{X}_2 + 2 \mathbb{X}_3, \nonumber                                                                          \\
		             & \qquad 4 \mathbb{X}_1 + \mathbb{X}_3 + \mathbb{X}_4, \mathbb{X}_1 + 3 \mathbb{X}_2 + \mathbb{X}_4 \},                                                                  \\
		\mathbb{V}_3 & = \{ 4 \mathbb{X}_1 + 3 \mathbb{X}_2 + 4 \mathbb{X}_3, 4 \mathbb{X}_2 + 2 \mathbb{X}_3 + 4 \mathbb{X}_4, \nonumber                                                     \\
		             & \qquad 4 \mathbb{X}_1 + 4 \mathbb{X}_3 + \mathbb{X}_4, 2 \mathbb{X}_1 + 3 \mathbb{X}_2 + 4 \mathbb{X}_4 \}. \label{main_eq_numerical_simulations_example_CRNs_004_002}
	\end{align}
\end{subequations}
The CRNs in Eq.~\eqref{main_eq_numerical_simulations_example_CRNs_004_001} belong to the class of CRNs labeled by $N_\mathbb{X} = 4$, $N_\mathbb{e} = 6$, and $m = 4$.
We call the CRNs in Eq.~\eqref{main_eq_numerical_simulations_example_CRNs_004_001} type 1, type 2, and type 3 from above.

\begin{figure}[t]
	\centering
	\includegraphics[scale=0.500]{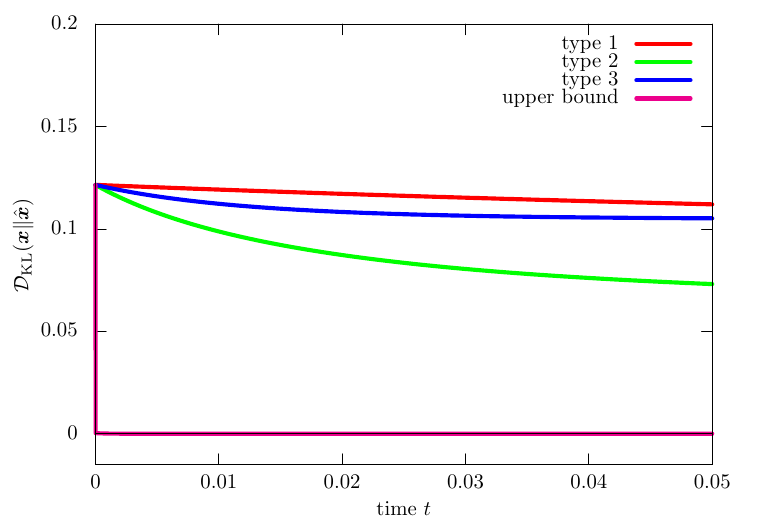}
	\caption{Dependence of $\mathcal{D}_\mathrm{KL} (\bm{x}_t \| \hat{\bm{x}})$ on time $t$ for the CRNs in Eq.~\eqref{main_eq_numerical_simulations_example_CRNs_004_001} and its upper bound in the case of $\bm{x}_\mathrm{eq} \ne \hat{\bm{x}}$.}
	\label{main_fig_upper_bound_not_eq_004_001}
\end{figure}

We plot the dynamics of the CRNs in Eq.~\eqref{main_eq_numerical_simulations_example_CRNs_004_001} and its upper bound in the case of $\bm{x}_\mathrm{eq} \ne \hat{\bm{x}}$.
In Fig.~\ref{main_fig_upper_bound_not_eq_004_001}, we set $\bm{x}_\mathrm{ini} = [9/8, 87/80, 27/20, 27/20]^\intercal$, $\hat{\bm{x}} = [1, 1, 1, 1]^\intercal$, $k_e^\pm = 1$, and $\Delta t = 1.0 \times 10^{-5}$.
Figure~\ref{main_fig_upper_bound_not_eq_004_001} clearly demonstrates the upper bound holds for $N_\mathbb{e} > 1$.

\begin{figure}[t]
	\centering
	\includegraphics[scale=0.500]{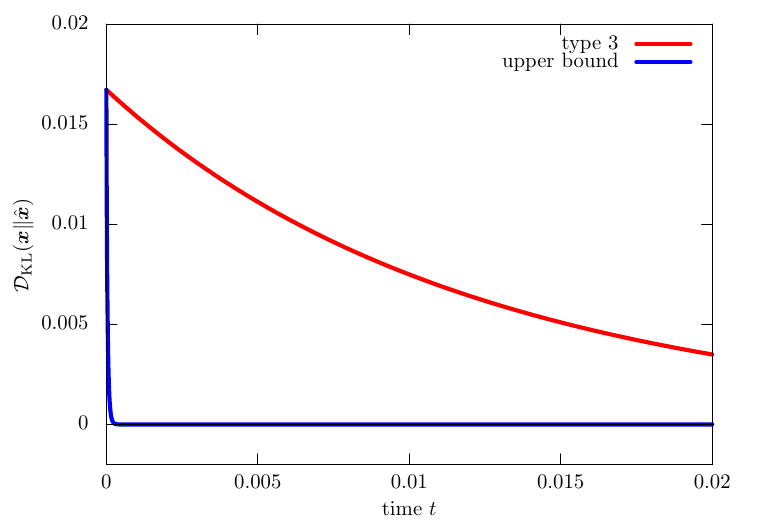}
	\caption{Dependence of $\mathcal{D}_\mathrm{KL} (\bm{x}_t \| \hat{\bm{x}})$ on time $t$ for the CRN in Eq.~\eqref{main_eq_numerical_simulations_example_CRNs_004_002} and its upper bound in the case of $\bm{x}_\mathrm{eq} = \hat{\bm{x}}$.}
	\label{main_fig_upper_bound_not_eq_004_002}
\end{figure}

\begin{figure}[t]
	\centering
	\includegraphics[scale=0.500]{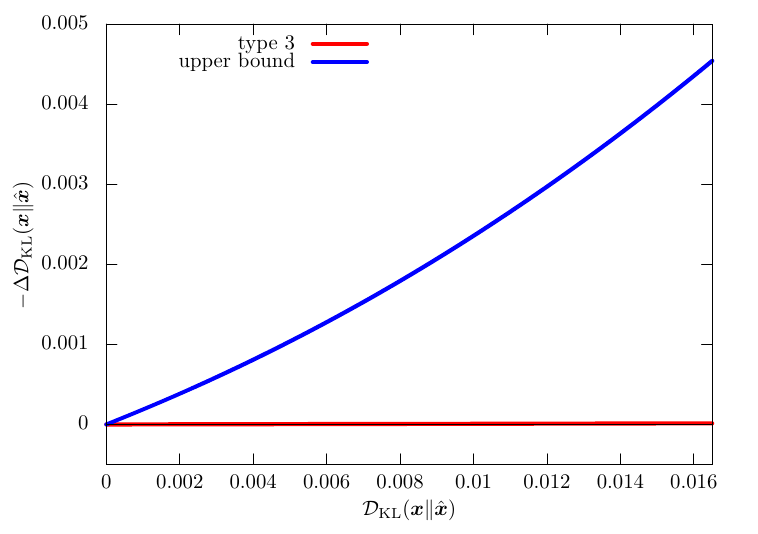}
	\caption{Dependence of $\mathcal{D}_\mathrm{KL} (\bm{x}_t \| \hat{\bm{x}})$ on time $t$ for the CRN in Eq.~\eqref{main_eq_numerical_simulations_example_CRNs_004_002} and its upper bound in the case of $\bm{x}_\mathrm{eq} = \hat{\bm{x}}$.}
	\label{main_fig_upper_bound_not_eq_004_003}
\end{figure}

We show the dependence of $\mathcal{D}_\mathrm{KL} (\bm{x}_t \| \hat{\bm{x}})$ on time $t$ for the CRN in Eq.~\eqref{main_eq_numerical_simulations_example_CRNs_004_002} and its upper bound in the case of $\bm{x}_\mathrm{eq} = \hat{\bm{x}}$.
In Fig.~\ref{main_fig_upper_bound_not_eq_004_002}, we set $\hat{\bm{x}} = [1.2547, 1.1021, 1.1951, 1.3388]^\intercal$.
In Fig.~\ref{main_fig_upper_bound_not_eq_004_003}, we also show the dependence of $\mathcal{D}_\mathrm{KL} (\bm{x}_t \| \hat{\bm{x}})$ on time $t$ for the CRN in Eq.~\eqref{main_eq_numerical_simulations_example_CRNs_004_002} and its upper bound in the case of $\bm{x}_\mathrm{eq} = \hat{\bm{x}}$.

\subsection{Comparison of the upper bounds}

\begin{figure}[t]
	\centering
	\includegraphics[scale=0.500]{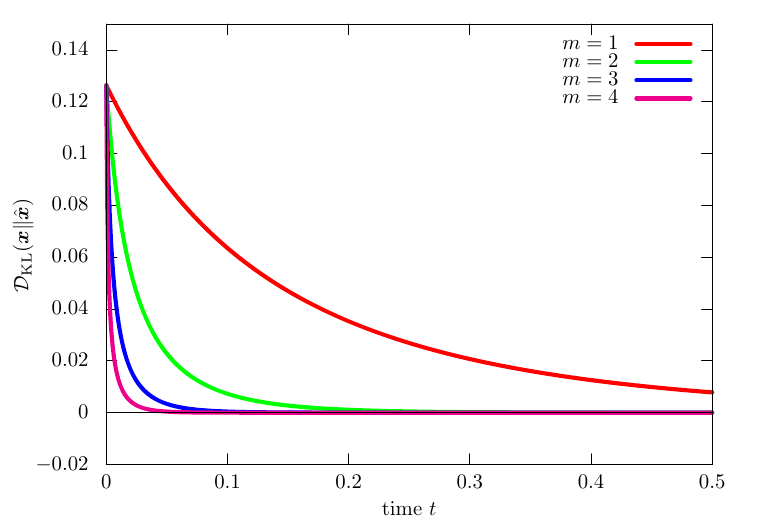}
	\caption{Dependence of $\mathcal{D}_\mathrm{KL} (\bm{x}_t \| \hat{\bm{x}})$ for $t$ for $m = 1, 2, 3, 4$. We set $N_\mathbb{X} = 4$ and $N_\mathbb{e} = 1$.}
	\label{main_fig_upper_bound_not_eq_005_001}
\end{figure}

\begin{figure}[t]
	\centering
	\includegraphics[scale=0.500]{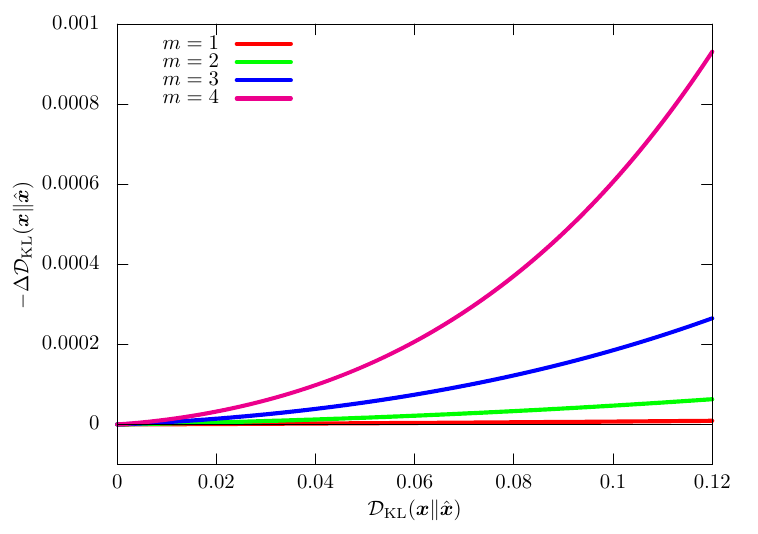}
	\caption{Relationship between $\mathcal{D}_\mathrm{KL} (\bm{x}_t \| \hat{\bm{x}})$ and $- \Delta \mathcal{D}_\mathrm{KL} (\bm{x}_t \| \hat{\bm{x}})$ for $N_\mathbb{X} = 4$ and $N_\mathbb{e} = 1$.}
	\label{main_fig_upper_bound_not_eq_005_002}
\end{figure}

In this section, we examine the behavior of the upper bound under varying parameters.
The parameters are \( N_\mathbb{X} = 4 \), \( N_\mathbb{e} = 1 \), \( \bm{x}_\mathrm{ini} = [3/4, 3/4, 5/4, 5/4]^\intercal \), and \( \bm{x}_\mathrm{eq} = [1, 1, 1, 1]^\intercal \).
Figure~\ref{main_fig_upper_bound_not_eq_005_001} depicts the dependence of \( \mathcal{D}_\mathrm{KL} (\bm{x}_t \| \hat{\bm{x}}) \) on \( t \) for \( m = 1, 2, 3, 4 \).
Figure~\ref{main_fig_upper_bound_not_eq_005_002} portrays the relationship between \( \mathcal{D}_\mathrm{KL} (\bm{x}_t \| \hat{\bm{x}}) \) and \( - \Delta \mathcal{D}_\mathrm{KL} (\bm{x}_t \| \hat{\bm{x}}) \) for \( N_\mathbb{X} = 4 \) and \( N_\mathbb{e} = 1 \).
The figures indicate that higher values of \( m \) are associated with increased rates of convergence.
This behavior is consistent with the expectation that nonlinearity in CRNs tends to influence reaction rates.

The relationship between the KL divergence and the entropy production was pointed out in Ref.~\cite{Sughiyama_002}.
Letting $\Sigma_\mathrm{tot} (\bm{x}_t)$ be the total entropy, the following relationship holds:
\begin{align}
	\Sigma_\mathrm{tot} (\bm{x}_t) - \Sigma_\mathrm{tot} (\bm{x}_{t'}) & = - \frac{V}{T} [\mathcal{D}_\mathrm{KL} (\bm{x}_t \| \hat{\bm{x}}) - \mathcal{D}_\mathrm{KL} (\bm{x}_{t'} \| \hat{\bm{x}})], \label{main_eq_relationship_entropy_production_divergence_001_001}
\end{align}
where $V$ is the volume of the system and $T$ is the temperature of the environment.
In NG, the right-hand side of Eq.~\eqref{main_eq_relationship_entropy_production_divergence_001_001} is maximized under $\mathcal{D}_{\phi'} (\bm{x}_t + \Delta \bm{x} \| \bm{x}_t) \le \epsilon$ as written in Eq.~\eqref{main_optimization_problem_NG_001_001}.
Furthermore, $\epsilon$ in the optimization problem to find the upper bound in Sec.~\ref{main_sec_upper_bound_001_001} is equal to or larger than the time-difference of the KL divergence of CRNs in a given class.
Thus, the entropy production of the system designed in Sec.~\ref{main_sec_upper_bound_001_001} is larger than those of CRNs in a given class and it shows faster convergence toward $\hat{\bm{x}}$.

\section{Discussion} \label{main_sec_discussions_001_001}

The relationship between the KL divergence and the entropy production was pointed out in Ref.~\cite{Sughiyama_002}.
Letting $\Sigma_\mathrm{tot} (\bm{x}_t)$ be the total entropy, the following relationship holds:
\begin{align}
	\Sigma_\mathrm{tot} (\bm{x}_t) - \Sigma_\mathrm{tot} (\bm{x}_{t'}) & = - \frac{V}{T} [\mathcal{D}_\mathrm{KL} (\bm{x}_t \| \hat{\bm{x}}) - \mathcal{D}_\mathrm{KL} (\bm{x}_{t'} \| \hat{\bm{x}})], \label{main_eq_relationship_entropy_production_divergence_001_001}
\end{align}
where $V$ is the volume of the system and $T$ is the temperature of the environment.
In NG, the right-hand side of Eq.~\eqref{main_eq_relationship_entropy_production_divergence_001_001} is maximized under $\mathcal{D}_{\phi'} (\bm{x}_t + \Delta \bm{x} \| \bm{x}_t) \le \epsilon$ as written in Eq.~\eqref{main_optimization_problem_NG_001_001}.
Furthermore, $\epsilon$ in the optimization problem to find the upper bound in Sec.~\ref{main_sec_upper_bound_001_001} is equal to or larger than the time-difference of the KL divergence of CRNs in a given class.
Thus, the entropy production of the system designed in Sec.~\ref{main_sec_upper_bound_001_001} is larger than those of CRNs in a given class and it shows faster convergence toward $\hat{\bm{x}}$.

\section{Conclusions} \label{main_sec_conclusions_001_001}

In this study, we developed a framework based on NG to establish an upper bound on the dynamics of a specific subset of CRNs.
The physical meaning of this bound relates to the concept of entropy production, which in turn is related to the speed of convergence of the chemical reaction.
The nonlinearity commonly present in CRNs presents a challenge, which is addressed here.
The optimization problem in the NG derivation was found to be related to entropy production, enriching the understanding of NG within a thermodynamic context.
While the primary focus has been on CRNs, the methods and discussions are applicable to a wider range of hypergraph dynamics.
The study holds implications for fields beyond chemistry and physics, including information science and machine learning.

\begin{acknowledgments}
	H.M. was supported by JSPS KAKENHI Grant Number 23H04489.
	T. J. K was supported by JST (Grants No. JPMJCR2011 and No. JPMJCR1927) and JSPS (Grant No. 19H05799).
	L.-S.B. was partially funded by  NSF award CHE-2002313.
\end{acknowledgments}

\appendix
\renewcommand{\theequation}{\Alph{section}.\arabic{equation}}

\section{Derivation of the KL divergence from the Bregman divergence} \label{main_sec_derivation_KL_divergence_001_001}

In this section, we show that the Bregman divergence, Eq.~\eqref{main_eq_def_Bregman_divergence_001_001}, with Eq.~\eqref{main_eq_potential_KL_001_001} is equivalent to the KL divergence, Eq.~\eqref{main_eq_def_KL_divergence_001_001}.
Let us define the following potential for $\alpha \in \mathbb{R}$:
\begin{align}
	\phi_\mathrm{KL}^{(\alpha)} (\bm{x}) & \coloneqq \sum_{i=1}^{N_\mathbb{X}} x_i (\ln x_i - \alpha) \label{main_eq_potential_KL_alpha_001_001}
\end{align}
The Bregman divergence, Eq.~\eqref{main_eq_def_Bregman_divergence_001_001}, with Eq.~\eqref{main_eq_potential_KL_alpha_001_001} is computed as follows:
\begin{align}
	\mathcal{D}_{\phi_\mathrm{KL}^{(\alpha)}} (\bm{x} \| \bm{y}) & = \phi_\mathrm{KL}^{(\alpha)} (\bm{x}) - \phi_\mathrm{KL}^{(\alpha)} (\bm{y}) - \langle (\bm{x} - \bm{y}), \nabla \phi_\mathrm{KL}^{(\alpha)} (\bm{y}) \rangle \\
	                                                             & = \sum_{i=1}^{N_\mathbb{X}} x_i (\ln x_i - \alpha) - \sum_{i=1}^{N_\mathbb{X}} y_i (\ln y_i - \alpha) \nonumber                                                \\
	                                                             & \quad - \sum_{i=1}^{N_\mathbb{X}} (x_i - y_i) (\ln y_i - \alpha + 1)                                                                                           \\
	                                                             & = \sum_{i=1}^{N_\mathbb{X}} x_i \ln x_i - \sum_{i=1}^{N_\mathbb{X}} y_i \ln y_i \nonumber                                                                      \\
	                                                             & \quad - \sum_{i=1}^{N_\mathbb{X}} (x_i - y_i) \ln y_i - \sum_{i=1}^{N_\mathbb{X}} (x_i - y_i)                                                                  \\
	                                                             & = \sum_{i=1}^{N_\mathbb{X}} x_i \ln \frac{x_i}{y_i} - \sum_{i=1}^{N_\mathbb{X}} (x_i - y_i)                                                                    \\
	                                                             & = \mathcal{D}_\mathrm{KL} (\bm{x} \| \bm{y})
\end{align}
Thus, the Bregman divergence, Eq.~\eqref{main_eq_def_Bregman_divergence_001_001}, with Eq.~\eqref{main_eq_potential_KL_alpha_001_001} is equivalent to the KL divergence, Eq.~\eqref{main_eq_def_KL_divergence_001_001}, independently from $\alpha$.
Furthermore, Eq.~\eqref{main_eq_potential_KL_001_001} is the special case of Eq.~\eqref{main_eq_potential_KL_alpha_001_001} with $\alpha = 0$.

\section{Upper bound system with the $L^2$ constraint} \label{main_sec_NG_L2_constraint_001_001}

In Sec.~\ref{main_sec_upper_bound_001_001}, we have considered $\phi_\mathrm{KL} (\cdot)$, Eq.~\eqref{main_eq_potential_KL_001_001}, as the potential of the Bregman divergence in the constraint term since the KL divergence is minimized in CRNs.
However, we are not limited to this choice, and it is expected that a different potential in the constraint may give us a different bound.
Another simple candidate for the potential of the Bregman divergence in the constraint is the $L^2$ norm given by
\begin{align}
	\phi_{L^2} (\bm{x}) & \coloneqq \sum_{i=1}^{N_\mathbb{X}} | x_i |^2.
\end{align}
In this case, $\mathcal{D}_\mathrm{KL} (\bm{x}_t + \| \dot{\bm{x}}_t \|_\mathrm{F} \bm{e}_t \Delta t \| \bm{x}_t)$ does not depend on $\bm{e}_t$ and the Hessian $G_{\phi_{L^2}} (\bm{x}_t)$ becomes the identity matrix: $G_{\phi_{L^2}} (\bm{x}_t) = \mathbb{1}$.

\bibliography{paper_speed-limit_CRN_999_001}

\begin{thebibliography}{36}%
\makeatletter
\providecommand \@ifxundefined [1]{%
 \@ifx{#1\undefined}
}%
\providecommand \@ifnum [1]{%
 \ifnum #1\expandafter \@firstoftwo
 \else \expandafter \@secondoftwo
 \fi
}%
\providecommand \@ifx [1]{%
 \ifx #1\expandafter \@firstoftwo
 \else \expandafter \@secondoftwo
 \fi
}%
\providecommand \natexlab [1]{#1}%
\providecommand \enquote  [1]{``#1''}%
\providecommand \bibnamefont  [1]{#1}%
\providecommand \bibfnamefont [1]{#1}%
\providecommand \citenamefont [1]{#1}%
\providecommand \href@noop [0]{\@secondoftwo}%
\providecommand \href [0]{\begingroup \@sanitize@url \@href}%
\providecommand \@href[1]{\@@startlink{#1}\@@href}%
\providecommand \@@href[1]{\endgroup#1\@@endlink}%
\providecommand \@sanitize@url [0]{\catcode `\\12\catcode `\$12\catcode
  `\&12\catcode `\#12\catcode `\^12\catcode `\_12\catcode `\%12\relax}%
\providecommand \@@startlink[1]{}%
\providecommand \@@endlink[0]{}%
\providecommand \url  [0]{\begingroup\@sanitize@url \@url }%
\providecommand \@url [1]{\endgroup\@href {#1}{\urlprefix }}%
\providecommand \urlprefix  [0]{URL }%
\providecommand \Eprint [0]{\href }%
\providecommand \doibase [0]{https://doi.org/}%
\providecommand \selectlanguage [0]{\@gobble}%
\providecommand \bibinfo  [0]{\@secondoftwo}%
\providecommand \bibfield  [0]{\@secondoftwo}%
\providecommand \translation [1]{[#1]}%
\providecommand \BibitemOpen [0]{}%
\providecommand \bibitemStop [0]{}%
\providecommand \bibitemNoStop [0]{.\EOS\space}%
\providecommand \EOS [0]{\spacefactor3000\relax}%
\providecommand \BibitemShut  [1]{\csname bibitem#1\endcsname}%
\let\auto@bib@innerbib\@empty
\bibitem [{\citenamefont {Seifert}(2012)}]{Seifert_001}%
  \BibitemOpen
  \bibfield  {author} {\bibinfo {author} {\bibfnamefont {U.}~\bibnamefont
  {Seifert}},\ }\bibfield  {title} {\bibinfo {title} {Stochastic
  thermodynamics, fluctuation theorems and molecular machines},\ }\href@noop {}
  {\bibfield  {journal} {\bibinfo  {journal} {Reports on progress in physics}\
  }\textbf {\bibinfo {volume} {75}},\ \bibinfo {pages} {126001} (\bibinfo
  {year} {2012})}\BibitemShut {NoStop}%
\bibitem [{\citenamefont {Shiraishi}(2023)}]{Shiraishi_001}%
  \BibitemOpen
  \bibfield  {author} {\bibinfo {author} {\bibfnamefont {N.}~\bibnamefont
  {Shiraishi}},\ }\href@noop {} {\emph {\bibinfo {title} {An Introduction to
  Stochastic Thermodynamics: From Basic to Advanced}}},\ Vol.\ \bibinfo
  {volume} {212}\ (\bibinfo  {publisher} {Springer Nature},\ \bibinfo {year}
  {2023})\BibitemShut {NoStop}%
\bibitem [{\citenamefont {Kawai}\ \emph {et~al.}(2007)\citenamefont {Kawai},
  \citenamefont {Parrondo},\ and\ \citenamefont {den Broeck}}]{Kawai_001}%
  \BibitemOpen
  \bibfield  {author} {\bibinfo {author} {\bibfnamefont {R.}~\bibnamefont
  {Kawai}}, \bibinfo {author} {\bibfnamefont {J.~M.~R.}\ \bibnamefont
  {Parrondo}},\ and\ \bibinfo {author} {\bibfnamefont {C.~V.}\ \bibnamefont
  {den Broeck}},\ }\bibfield  {title} {\bibinfo {title} {Dissipation: The
  phase-space perspective},\ }\href
  {https://doi.org/10.1103/PhysRevLett.98.080602} {\bibfield  {journal}
  {\bibinfo  {journal} {Phys. Rev. Lett.}\ }\textbf {\bibinfo {volume} {98}},\
  \bibinfo {pages} {080602} (\bibinfo {year} {2007})}\BibitemShut {NoStop}%
\bibitem [{\citenamefont {Miyahara}\ and\ \citenamefont
  {Aihara}(2018)}]{Miyahara_001}%
  \BibitemOpen
  \bibfield  {author} {\bibinfo {author} {\bibfnamefont {H.}~\bibnamefont
  {Miyahara}}\ and\ \bibinfo {author} {\bibfnamefont {K.}~\bibnamefont
  {Aihara}},\ }\bibfield  {title} {\bibinfo {title} {Work relations with
  measurement and feedback control on nonuniform temperature systems},\ }\href
  {https://doi.org/10.1103/PhysRevE.98.042138} {\bibfield  {journal} {\bibinfo
  {journal} {Phys. Rev. E}\ }\textbf {\bibinfo {volume} {98}},\ \bibinfo
  {pages} {042138} (\bibinfo {year} {2018})}\BibitemShut {NoStop}%
\bibitem [{\citenamefont {Barato}\ and\ \citenamefont
  {Seifert}(2015)}]{Barato_001}%
  \BibitemOpen
  \bibfield  {author} {\bibinfo {author} {\bibfnamefont {A.~C.}\ \bibnamefont
  {Barato}}\ and\ \bibinfo {author} {\bibfnamefont {U.}~\bibnamefont
  {Seifert}},\ }\bibfield  {title} {\bibinfo {title} {Thermodynamic uncertainty
  relation for biomolecular processes},\ }\href
  {https://doi.org/10.1103/PhysRevLett.114.158101} {\bibfield  {journal}
  {\bibinfo  {journal} {Phys. Rev. Lett.}\ }\textbf {\bibinfo {volume} {114}},\
  \bibinfo {pages} {158101} (\bibinfo {year} {2015})}\BibitemShut {NoStop}%
\bibitem [{\citenamefont {Gingrich}\ \emph {et~al.}(2016)\citenamefont
  {Gingrich}, \citenamefont {Horowitz}, \citenamefont {Perunov},\ and\
  \citenamefont {England}}]{Gingrich_001}%
  \BibitemOpen
  \bibfield  {author} {\bibinfo {author} {\bibfnamefont {T.~R.}\ \bibnamefont
  {Gingrich}}, \bibinfo {author} {\bibfnamefont {J.~M.}\ \bibnamefont
  {Horowitz}}, \bibinfo {author} {\bibfnamefont {N.}~\bibnamefont {Perunov}},\
  and\ \bibinfo {author} {\bibfnamefont {J.~L.}\ \bibnamefont {England}},\
  }\bibfield  {title} {\bibinfo {title} {Dissipation bounds all steady-state
  current fluctuations},\ }\href
  {https://doi.org/10.1103/PhysRevLett.116.120601} {\bibfield  {journal}
  {\bibinfo  {journal} {Phys. Rev. Lett.}\ }\textbf {\bibinfo {volume} {116}},\
  \bibinfo {pages} {120601} (\bibinfo {year} {2016})}\BibitemShut {NoStop}%
\bibitem [{\citenamefont {Hasegawa}\ and\ \citenamefont
  {Van~Vu}(2019)}]{Hasegawa_001}%
  \BibitemOpen
  \bibfield  {author} {\bibinfo {author} {\bibfnamefont {Y.}~\bibnamefont
  {Hasegawa}}\ and\ \bibinfo {author} {\bibfnamefont {T.}~\bibnamefont
  {Van~Vu}},\ }\bibfield  {title} {\bibinfo {title} {Uncertainty relations in
  stochastic processes: An information inequality approach},\ }\href
  {https://doi.org/10.1103/PhysRevE.99.062126} {\bibfield  {journal} {\bibinfo
  {journal} {Phys. Rev. E}\ }\textbf {\bibinfo {volume} {99}},\ \bibinfo
  {pages} {062126} (\bibinfo {year} {2019})}\BibitemShut {NoStop}%
\bibitem [{\citenamefont {Van~Vu}\ and\ \citenamefont
  {Hasegawa}(2019)}]{Vu_001}%
  \BibitemOpen
  \bibfield  {author} {\bibinfo {author} {\bibfnamefont {T.}~\bibnamefont
  {Van~Vu}}\ and\ \bibinfo {author} {\bibfnamefont {Y.}~\bibnamefont
  {Hasegawa}},\ }\bibfield  {title} {\bibinfo {title} {Uncertainty relations
  for underdamped langevin dynamics},\ }\href
  {https://doi.org/10.1103/PhysRevE.100.032130} {\bibfield  {journal} {\bibinfo
   {journal} {Phys. Rev. E}\ }\textbf {\bibinfo {volume} {100}},\ \bibinfo
  {pages} {032130} (\bibinfo {year} {2019})}\BibitemShut {NoStop}%
\bibitem [{\citenamefont {Van~Vu}\ and\ \citenamefont {Saito}(2023)}]{Vu_002}%
  \BibitemOpen
  \bibfield  {author} {\bibinfo {author} {\bibfnamefont {T.}~\bibnamefont
  {Van~Vu}}\ and\ \bibinfo {author} {\bibfnamefont {K.}~\bibnamefont {Saito}},\
  }\bibfield  {title} {\bibinfo {title} {Thermodynamic unification of optimal
  transport: Thermodynamic uncertainty relation, minimum dissipation, and
  thermodynamic speed limits},\ }\href
  {https://doi.org/10.1103/PhysRevX.13.011013} {\bibfield  {journal} {\bibinfo
  {journal} {Phys. Rev. X}\ }\textbf {\bibinfo {volume} {13}},\ \bibinfo
  {pages} {011013} (\bibinfo {year} {2023})}\BibitemShut {NoStop}%
\bibitem [{\citenamefont {Ohga}\ \emph {et~al.}(2023)\citenamefont {Ohga},
  \citenamefont {Ito},\ and\ \citenamefont {Kolchinsky}}]{Ohga_001}%
  \BibitemOpen
  \bibfield  {author} {\bibinfo {author} {\bibfnamefont {N.}~\bibnamefont
  {Ohga}}, \bibinfo {author} {\bibfnamefont {S.}~\bibnamefont {Ito}},\ and\
  \bibinfo {author} {\bibfnamefont {A.}~\bibnamefont {Kolchinsky}},\ }\bibfield
   {title} {\bibinfo {title} {Thermodynamic bound on the asymmetry of
  cross-correlations},\ }\href@noop {} {\bibfield  {journal} {\bibinfo
  {journal} {arXiv preprint arXiv:2303.13116}\ } (\bibinfo {year}
  {2023})}\BibitemShut {NoStop}%
\bibitem [{\citenamefont {Van~Vu}\ \emph {et~al.}(2023)\citenamefont {Van~Vu},
  \citenamefont {Vo},\ and\ \citenamefont {Saito}}]{Vu_003}%
  \BibitemOpen
  \bibfield  {author} {\bibinfo {author} {\bibfnamefont {T.}~\bibnamefont
  {Van~Vu}}, \bibinfo {author} {\bibfnamefont {V.~T.}\ \bibnamefont {Vo}},\
  and\ \bibinfo {author} {\bibfnamefont {K.}~\bibnamefont {Saito}},\ }\bibfield
   {title} {\bibinfo {title} {Dissipation bounds asymmetry of finite-time
  cross-correlations},\ }\href@noop {} {\bibfield  {journal} {\bibinfo
  {journal} {arXiv preprint arXiv:2305.18000}\ } (\bibinfo {year}
  {2023})}\BibitemShut {NoStop}%
\bibitem [{\citenamefont {Dechant}\ \emph {et~al.}(2023)\citenamefont
  {Dechant}, \citenamefont {Garnier-Brun},\ and\ \citenamefont
  {Sasa}}]{Dechant_001}%
  \BibitemOpen
  \bibfield  {author} {\bibinfo {author} {\bibfnamefont {A.}~\bibnamefont
  {Dechant}}, \bibinfo {author} {\bibfnamefont {J.}~\bibnamefont
  {Garnier-Brun}},\ and\ \bibinfo {author} {\bibfnamefont {S.-i.}\ \bibnamefont
  {Sasa}},\ }\bibfield  {title} {\bibinfo {title} {Thermodynamic bounds on
  correlation times},\ }\href@noop {} {\bibfield  {journal} {\bibinfo
  {journal} {arXiv preprint arXiv:2303.13038}\ } (\bibinfo {year}
  {2023})}\BibitemShut {NoStop}%
\bibitem [{\citenamefont {Kobayashi}\ and\ \citenamefont
  {Sughiyama}(2015)}]{Kobayashi_001}%
  \BibitemOpen
  \bibfield  {author} {\bibinfo {author} {\bibfnamefont {T.~J.}\ \bibnamefont
  {Kobayashi}}\ and\ \bibinfo {author} {\bibfnamefont {Y.}~\bibnamefont
  {Sughiyama}},\ }\bibfield  {title} {\bibinfo {title} {Fluctuation relations
  of fitness and information in population dynamics},\ }\href
  {https://doi.org/10.1103/PhysRevLett.115.238102} {\bibfield  {journal}
  {\bibinfo  {journal} {Phys. Rev. Lett.}\ }\textbf {\bibinfo {volume} {115}},\
  \bibinfo {pages} {238102} (\bibinfo {year} {2015})}\BibitemShut {NoStop}%
\bibitem [{\citenamefont {Miyahara}(2019)}]{Miyahara_002}%
  \BibitemOpen
  \bibfield  {author} {\bibinfo {author} {\bibfnamefont {H.}~\bibnamefont
  {Miyahara}},\ }\bibfield  {title} {\bibinfo {title} {Many-body perturbation
  theory and fluctuation relations for interacting population dynamics},\
  }\href {https://doi.org/10.1103/PhysRevE.99.042415} {\bibfield  {journal}
  {\bibinfo  {journal} {Phys. Rev. E}\ }\textbf {\bibinfo {volume} {99}},\
  \bibinfo {pages} {042415} (\bibinfo {year} {2019})}\BibitemShut {NoStop}%
\bibitem [{\citenamefont {Sughiyama}\ and\ \citenamefont
  {Kobayashi}(2017)}]{Sughiyama_001}%
  \BibitemOpen
  \bibfield  {author} {\bibinfo {author} {\bibfnamefont {Y.}~\bibnamefont
  {Sughiyama}}\ and\ \bibinfo {author} {\bibfnamefont {T.~J.}\ \bibnamefont
  {Kobayashi}},\ }\bibfield  {title} {\bibinfo {title} {Steady-state
  thermodynamics for population growth in fluctuating environments},\ }\href
  {https://doi.org/10.1103/PhysRevE.95.012131} {\bibfield  {journal} {\bibinfo
  {journal} {Phys. Rev. E}\ }\textbf {\bibinfo {volume} {95}},\ \bibinfo
  {pages} {012131} (\bibinfo {year} {2017})}\BibitemShut {NoStop}%
\bibitem [{\citenamefont {Miyahara}(2022)}]{Miyahara_003}%
  \BibitemOpen
  \bibfield  {author} {\bibinfo {author} {\bibfnamefont {H.}~\bibnamefont
  {Miyahara}},\ }\bibfield  {title} {\bibinfo {title} {Steady-state
  thermodynamics for population dynamics in fluctuating environments with side
  information},\ }\href {https://doi.org/10.1088/1742-5468/ac42cc} {\bibfield
  {journal} {\bibinfo  {journal} {Journal of Statistical Mechanics: Theory and
  Experiment}\ }\textbf {\bibinfo {volume} {2022}},\ \bibinfo {pages} {013501}
  (\bibinfo {year} {2022})}\BibitemShut {NoStop}%
\bibitem [{\citenamefont {Adachi}\ \emph {et~al.}(2022)\citenamefont {Adachi},
  \citenamefont {Iritani},\ and\ \citenamefont {Hamazaki}}]{Adachi_001}%
  \BibitemOpen
  \bibfield  {author} {\bibinfo {author} {\bibfnamefont {K.}~\bibnamefont
  {Adachi}}, \bibinfo {author} {\bibfnamefont {R.}~\bibnamefont {Iritani}},\
  and\ \bibinfo {author} {\bibfnamefont {R.}~\bibnamefont {Hamazaki}},\
  }\bibfield  {title} {\bibinfo {title} {Universal constraint on nonlinear
  population dynamics},\ }\href@noop {} {\bibfield  {journal} {\bibinfo
  {journal} {Communications Physics}\ }\textbf {\bibinfo {volume} {5}},\
  \bibinfo {pages} {129} (\bibinfo {year} {2022})}\BibitemShut {NoStop}%
\bibitem [{\citenamefont {Hoshino}\ \emph {et~al.}(2023)\citenamefont
  {Hoshino}, \citenamefont {Nagayama}, \citenamefont {Yoshimura}, \citenamefont
  {Yamagishi},\ and\ \citenamefont {Ito}}]{Hoshino_001}%
  \BibitemOpen
  \bibfield  {author} {\bibinfo {author} {\bibfnamefont {M.}~\bibnamefont
  {Hoshino}}, \bibinfo {author} {\bibfnamefont {R.}~\bibnamefont {Nagayama}},
  \bibinfo {author} {\bibfnamefont {K.}~\bibnamefont {Yoshimura}}, \bibinfo
  {author} {\bibfnamefont {J.~F.}\ \bibnamefont {Yamagishi}},\ and\ \bibinfo
  {author} {\bibfnamefont {S.}~\bibnamefont {Ito}},\ }\bibfield  {title}
  {\bibinfo {title} {Geometric speed limit for acceleration by natural
  selection in evolutionary processes},\ }\href
  {https://doi.org/10.1103/PhysRevResearch.5.023127} {\bibfield  {journal}
  {\bibinfo  {journal} {Phys. Rev. Res.}\ }\textbf {\bibinfo {volume} {5}},\
  \bibinfo {pages} {023127} (\bibinfo {year} {2023})}\BibitemShut {NoStop}%
\bibitem [{\citenamefont {Kobayashi}\ \emph
  {et~al.}(2022{\natexlab{a}})\citenamefont {Kobayashi}, \citenamefont
  {Loutchko}, \citenamefont {Kamimura}, \citenamefont {Horiguchi},\ and\
  \citenamefont {Sughiyama}}]{Kobayashi_003}%
  \BibitemOpen
  \bibfield  {author} {\bibinfo {author} {\bibfnamefont {T.~J.}\ \bibnamefont
  {Kobayashi}}, \bibinfo {author} {\bibfnamefont {D.}~\bibnamefont {Loutchko}},
  \bibinfo {author} {\bibfnamefont {A.}~\bibnamefont {Kamimura}}, \bibinfo
  {author} {\bibfnamefont {S.}~\bibnamefont {Horiguchi}},\ and\ \bibinfo
  {author} {\bibfnamefont {Y.}~\bibnamefont {Sughiyama}},\ }\bibfield  {title}
  {\bibinfo {title} {Information geometry of dynamics on graphs and
  hypergraphs},\ }\href@noop {} {\bibfield  {journal} {\bibinfo  {journal}
  {arXiv preprint arXiv:2211.14455}\ } (\bibinfo {year}
  {2022}{\natexlab{a}})}\BibitemShut {NoStop}%
\bibitem [{\citenamefont {Kobayashi}\ \emph
  {et~al.}(2022{\natexlab{b}})\citenamefont {Kobayashi}, \citenamefont
  {Loutchko}, \citenamefont {Kamimura},\ and\ \citenamefont
  {Sughiyama}}]{Kobayashi_004}%
  \BibitemOpen
  \bibfield  {author} {\bibinfo {author} {\bibfnamefont {T.~J.}\ \bibnamefont
  {Kobayashi}}, \bibinfo {author} {\bibfnamefont {D.}~\bibnamefont {Loutchko}},
  \bibinfo {author} {\bibfnamefont {A.}~\bibnamefont {Kamimura}},\ and\
  \bibinfo {author} {\bibfnamefont {Y.}~\bibnamefont {Sughiyama}},\ }\bibfield
  {title} {\bibinfo {title} {Kinetic derivation of the hessian geometric
  structure in chemical reaction networks},\ }\href
  {https://doi.org/10.1103/PhysRevResearch.4.033066} {\bibfield  {journal}
  {\bibinfo  {journal} {Phys. Rev. Res.}\ }\textbf {\bibinfo {volume} {4}},\
  \bibinfo {pages} {033066} (\bibinfo {year} {2022}{\natexlab{b}})}\BibitemShut
  {NoStop}%
\bibitem [{\citenamefont {Kobayashi}\ \emph
  {et~al.}(2022{\natexlab{c}})\citenamefont {Kobayashi}, \citenamefont
  {Loutchko}, \citenamefont {Kamimura},\ and\ \citenamefont
  {Sughiyama}}]{Kobayashi_005}%
  \BibitemOpen
  \bibfield  {author} {\bibinfo {author} {\bibfnamefont {T.~J.}\ \bibnamefont
  {Kobayashi}}, \bibinfo {author} {\bibfnamefont {D.}~\bibnamefont {Loutchko}},
  \bibinfo {author} {\bibfnamefont {A.}~\bibnamefont {Kamimura}},\ and\
  \bibinfo {author} {\bibfnamefont {Y.}~\bibnamefont {Sughiyama}},\ }\bibfield
  {title} {\bibinfo {title} {Hessian geometry of nonequilibrium chemical
  reaction networks and entropy production decompositions},\ }\href
  {https://doi.org/10.1103/PhysRevResearch.4.033208} {\bibfield  {journal}
  {\bibinfo  {journal} {Phys. Rev. Res.}\ }\textbf {\bibinfo {volume} {4}},\
  \bibinfo {pages} {033208} (\bibinfo {year} {2022}{\natexlab{c}})}\BibitemShut
  {NoStop}%
\bibitem [{\citenamefont {Okada}\ and\ \citenamefont
  {Mochizuki}(2016)}]{Okada_001}%
  \BibitemOpen
  \bibfield  {author} {\bibinfo {author} {\bibfnamefont {T.}~\bibnamefont
  {Okada}}\ and\ \bibinfo {author} {\bibfnamefont {A.}~\bibnamefont
  {Mochizuki}},\ }\bibfield  {title} {\bibinfo {title} {Law of localization in
  chemical reaction networks},\ }\href
  {https://doi.org/10.1103/PhysRevLett.117.048101} {\bibfield  {journal}
  {\bibinfo  {journal} {Phys. Rev. Lett.}\ }\textbf {\bibinfo {volume} {117}},\
  \bibinfo {pages} {048101} (\bibinfo {year} {2016})}\BibitemShut {NoStop}%
\bibitem [{\citenamefont {Hirono}\ \emph {et~al.}(2021)\citenamefont {Hirono},
  \citenamefont {Okada}, \citenamefont {Miyazaki},\ and\ \citenamefont
  {Hidaka}}]{Hirono_001}%
  \BibitemOpen
  \bibfield  {author} {\bibinfo {author} {\bibfnamefont {Y.}~\bibnamefont
  {Hirono}}, \bibinfo {author} {\bibfnamefont {T.}~\bibnamefont {Okada}},
  \bibinfo {author} {\bibfnamefont {H.}~\bibnamefont {Miyazaki}},\ and\
  \bibinfo {author} {\bibfnamefont {Y.}~\bibnamefont {Hidaka}},\ }\bibfield
  {title} {\bibinfo {title} {Structural reduction of chemical reaction networks
  based on topology},\ }\href
  {https://doi.org/10.1103/PhysRevResearch.3.043123} {\bibfield  {journal}
  {\bibinfo  {journal} {Phys. Rev. Res.}\ }\textbf {\bibinfo {volume} {3}},\
  \bibinfo {pages} {043123} (\bibinfo {year} {2021})}\BibitemShut {NoStop}%
\bibitem [{\citenamefont {Hirono}\ \emph {et~al.}()\citenamefont {Hirono},
  \citenamefont {Okada}, \citenamefont {Miyazaki},\ and\ \citenamefont
  {Hidaka}}]{Hirono_002}%
  \BibitemOpen
  \bibfield  {author} {\bibinfo {author} {\bibfnamefont {Y.}~\bibnamefont
  {Hirono}}, \bibinfo {author} {\bibfnamefont {T.}~\bibnamefont {Okada}},
  \bibinfo {author} {\bibfnamefont {H.}~\bibnamefont {Miyazaki}},\ and\
  \bibinfo {author} {\bibfnamefont {Y.}~\bibnamefont {Hidaka}},\ }\bibinfo
  {title} {Structural reduction of hypergraphs based on topology},\ in\ \href
  {https://doi.org/10.7566/JPSCP.36.011008} {\emph {\bibinfo {booktitle}
  {Proceedings of Blockchain in Kyoto 2021 (BCK21)}}},\ \Eprint
  {https://arxiv.org/abs/https://journals.jps.jp/doi/pdf/10.7566/JPSCP.36.011008}
  {https://journals.jps.jp/doi/pdf/10.7566/JPSCP.36.011008} \BibitemShut
  {NoStop}%
\bibitem [{\citenamefont {Hirono}\ \emph {et~al.}(2023)\citenamefont {Hirono},
  \citenamefont {Hong},\ and\ \citenamefont {Kim}}]{Hirono_003}%
  \BibitemOpen
  \bibfield  {author} {\bibinfo {author} {\bibfnamefont {Y.}~\bibnamefont
  {Hirono}}, \bibinfo {author} {\bibfnamefont {H.}~\bibnamefont {Hong}},\ and\
  \bibinfo {author} {\bibfnamefont {J.~K.}\ \bibnamefont {Kim}},\ }\bibfield
  {title} {\bibinfo {title} {Robust perfect adaptation of reaction fluxes
  ensured by network topology},\ }\href@noop {} {\bibfield  {journal} {\bibinfo
   {journal} {arXiv preprint arXiv:2302.01270}\ } (\bibinfo {year}
  {2023})}\BibitemShut {NoStop}%
\bibitem [{\citenamefont {Amari}\ and\ \citenamefont
  {Nagaoka}(2000)}]{Amari_002}%
  \BibitemOpen
  \bibfield  {author} {\bibinfo {author} {\bibfnamefont {S.-i.}\ \bibnamefont
  {Amari}}\ and\ \bibinfo {author} {\bibfnamefont {H.}~\bibnamefont
  {Nagaoka}},\ }\href@noop {} {\emph {\bibinfo {title} {Methods of information
  geometry}}},\ Vol.\ \bibinfo {volume} {191}\ (\bibinfo  {publisher} {American
  Mathematical Soc.},\ \bibinfo {year} {2000})\BibitemShut {NoStop}%
\bibitem [{\citenamefont {Amari}(2016)}]{Amari_004}%
  \BibitemOpen
  \bibfield  {author} {\bibinfo {author} {\bibfnamefont {S.-i.}\ \bibnamefont
  {Amari}},\ }\href {https://books.google.co.jp/books?id=UkSFCwAAQBAJ} {\emph
  {\bibinfo {title} {Information Geometry and Its Applications}}},\ Applied
  Mathematical Sciences\ (\bibinfo  {publisher} {Springer Japan},\ \bibinfo
  {year} {2016})\BibitemShut {NoStop}%
\bibitem [{\citenamefont {Amari}(1998)}]{Amari_005}%
  \BibitemOpen
  \bibfield  {author} {\bibinfo {author} {\bibfnamefont {S.-I.}\ \bibnamefont
  {Amari}},\ }\bibfield  {title} {\bibinfo {title} {Natural gradient works
  efficiently in learning},\ }\href@noop {} {\bibfield  {journal} {\bibinfo
  {journal} {Neural computation}\ }\textbf {\bibinfo {volume} {10}},\ \bibinfo
  {pages} {251} (\bibinfo {year} {1998})}\BibitemShut {NoStop}%
\bibitem [{\citenamefont {Wang}\ and\ \citenamefont {Li}(2022)}]{Wang_001}%
  \BibitemOpen
  \bibfield  {author} {\bibinfo {author} {\bibfnamefont {Y.}~\bibnamefont
  {Wang}}\ and\ \bibinfo {author} {\bibfnamefont {W.}~\bibnamefont {Li}},\
  }\bibfield  {title} {\bibinfo {title} {Accelerated information gradient
  flow},\ }\href@noop {} {\bibfield  {journal} {\bibinfo  {journal} {Journal of
  Scientific Computing}\ }\textbf {\bibinfo {volume} {90}},\ \bibinfo {pages}
  {1} (\bibinfo {year} {2022})}\BibitemShut {NoStop}%
\bibitem [{\citenamefont {Horiguchi}\ and\ \citenamefont
  {Kobayashi}(2023)}]{Horiguchi_001}%
  \BibitemOpen
  \bibfield  {author} {\bibinfo {author} {\bibfnamefont {S.~A.}\ \bibnamefont
  {Horiguchi}}\ and\ \bibinfo {author} {\bibfnamefont {T.~J.}\ \bibnamefont
  {Kobayashi}},\ }\bibfield  {title} {\bibinfo {title} {Cellular gradient flow
  structure linking single-cell-level rules and population-level dynamics},\
  }\href {https://doi.org/10.1103/PhysRevResearch.5.L022052} {\bibfield
  {journal} {\bibinfo  {journal} {Phys. Rev. Res.}\ }\textbf {\bibinfo {volume}
  {5}},\ \bibinfo {pages} {L022052} (\bibinfo {year} {2023})}\BibitemShut
  {NoStop}%
\bibitem [{\citenamefont {Karbowski}(2023)}]{Karbowski_001}%
  \BibitemOpen
  \bibfield  {author} {\bibinfo {author} {\bibfnamefont {J.}~\bibnamefont
  {Karbowski}},\ }\bibfield  {title} {\bibinfo {title} {Bounds on the rates of
  statistical divergences and mutual information},\ }\href@noop {} {\bibfield
  {journal} {\bibinfo  {journal} {arXiv preprint arXiv:2308.05597}\ } (\bibinfo
  {year} {2023})}\BibitemShut {NoStop}%
\bibitem [{\citenamefont {Beard}\ and\ \citenamefont {Qian}(2008)}]{Beard_001}%
  \BibitemOpen
  \bibfield  {author} {\bibinfo {author} {\bibfnamefont {D.~A.}\ \bibnamefont
  {Beard}}\ and\ \bibinfo {author} {\bibfnamefont {H.}~\bibnamefont {Qian}},\
  }\href@noop {} {\emph {\bibinfo {title} {Chemical biophysics: quantitative
  analysis of cellular systems}}},\ Vol.\ \bibinfo {volume} {126}\ (\bibinfo
  {publisher} {Cambridge University Press Cambridge},\ \bibinfo {year}
  {2008})\BibitemShut {NoStop}%
\bibitem [{\citenamefont {Feinberg}(2019)}]{Feinberg_001}%
  \BibitemOpen
  \bibfield  {author} {\bibinfo {author} {\bibfnamefont {M.}~\bibnamefont
  {Feinberg}},\ }\href@noop {} {\emph {\bibinfo {title} {Foundations of
  chemical reaction network theory}}}\ (\bibinfo  {publisher} {Springer},\
  \bibinfo {year} {2019})\BibitemShut {NoStop}%
\bibitem [{\citenamefont {Rao}\ and\ \citenamefont {Esposito}(2016)}]{Rao_001}%
  \BibitemOpen
  \bibfield  {author} {\bibinfo {author} {\bibfnamefont {R.}~\bibnamefont
  {Rao}}\ and\ \bibinfo {author} {\bibfnamefont {M.}~\bibnamefont {Esposito}},\
  }\bibfield  {title} {\bibinfo {title} {Nonequilibrium thermodynamics of
  chemical reaction networks: Wisdom from stochastic thermodynamics},\ }\href
  {https://doi.org/10.1103/PhysRevX.6.041064} {\bibfield  {journal} {\bibinfo
  {journal} {Phys. Rev. X}\ }\textbf {\bibinfo {volume} {6}},\ \bibinfo {pages}
  {041064} (\bibinfo {year} {2016})}\BibitemShut {NoStop}%
\bibitem [{\citenamefont {Hamming}(2012)}]{Hamming_001}%
  \BibitemOpen
  \bibfield  {author} {\bibinfo {author} {\bibfnamefont {R.}~\bibnamefont
  {Hamming}},\ }\href@noop {} {\emph {\bibinfo {title} {Numerical methods for
  scientists and engineers}}}\ (\bibinfo  {publisher} {Courier Corporation},\
  \bibinfo {year} {2012})\BibitemShut {NoStop}%
\bibitem [{\citenamefont {Sughiyama}\ \emph {et~al.}(2022)\citenamefont
  {Sughiyama}, \citenamefont {Loutchko}, \citenamefont {Kamimura},\ and\
  \citenamefont {Kobayashi}}]{Sughiyama_002}%
  \BibitemOpen
  \bibfield  {author} {\bibinfo {author} {\bibfnamefont {Y.}~\bibnamefont
  {Sughiyama}}, \bibinfo {author} {\bibfnamefont {D.}~\bibnamefont {Loutchko}},
  \bibinfo {author} {\bibfnamefont {A.}~\bibnamefont {Kamimura}},\ and\
  \bibinfo {author} {\bibfnamefont {T.~J.}\ \bibnamefont {Kobayashi}},\
  }\bibfield  {title} {\bibinfo {title} {Hessian geometric structure of
  chemical thermodynamic systems with stoichiometric constraints},\ }\href
  {https://doi.org/10.1103/PhysRevResearch.4.033065} {\bibfield  {journal}
  {\bibinfo  {journal} {Phys. Rev. Res.}\ }\textbf {\bibinfo {volume} {4}},\
  \bibinfo {pages} {033065} (\bibinfo {year} {2022})}\BibitemShut {NoStop}%
\end{thebibliography}%

\end{document}